\documentstyle[12pt,epsf]{article}

\begin{document}
\begin{flushright}
SLAC--PUB-7890\\
August 1998\\  
\end{flushright}

\bigskip\bigskip
\begin{center}

{{\bf\large  PROCESS, SYSTEM, CAUSALITY, AND QUANTUM MECHANICS}\\
{\it {\bf A Psychoanalysis of Animal Faith}}\footnote{\baselineskip=12pt
Work supported in part by Department of Energy contract DE--AC03--76SF00515.}}

\bigskip

Tom Etter \\
112 Blackburn Avenue\\ 
Menlo Park, California 94025--2704 \\

and\\
H. Pierre Noyes \\
Stanford Linear Accelerator Center\\
Stanford University, Stanford, CA 94309\\
\end{center}
\vfill
\begin{center}
Submitted to {\it International Journal of Theoretical Physics.}
\end{center}
\vfill

\newpage

\vfill
\begin{abstract}
We shall argue in this paper that a central piece of
modern physics does not really belong to physics at all but to
elementary probability theory.  Given a joint probability distribution J
on a set of random variables containing x and y, define a link between x
and y to be the condition x=y on J. Define the {\it state} D of a link x=y as
the joint probability distribution matrix on x and y without the link.
The two core laws of quantum mechanics are the Born probability rule,
and the unitary dynamical law whose best known form is the Schrodinger's
equation.  Von Neumann formulated these two laws in the language of
Hilbert space as prob(P) = trace(PD) and D'T = TD respectively, where P
is a projection, D and D' are (von Neumann) density matrices, and T is a unitary
transformation.  We'll see that if we regard link states as density
matrices, the algebraic forms of these two core laws occur as completely
general theorems about links.  When we extend probability theory by
allowing cases to count negatively, we find that the Hilbert space
framework of quantum mechanics proper emerges from the assumption that
all D's are symmetrical in rows and columns.  On the other hand,
Markovian systems emerge when we assume that one of every linked
variable pair has a uniform probability distribution.  By representing
quantum and Markovian structure in this way, we see clearly both how
they differ, and also how they can coexist in natural harmony with each
other, as they must in quantum measurement, which we'll examine in some
detail.  Looking beyond quantum mechanics, we see how both structures
have their special places in a much larger continuum of formal systems
that we have yet to look for in nature.

\end{abstract}

\begin{center}
{\bf INTRODUCTION: COUNTING SHEEP}
\end{center}

Once upon a time there was a sheep farmer who had ten small barns, in
each of which he kept five sheep.  When asked how many sheep he had
altogether, he replied ``many", for people in those days counted on their
fingers, and no one had ever thought of counting beyond ten.

Every morning he would drive his sheep over the hill and through the
woods to their pasture, where they assembled in five fields, ten sheep
to a field.  The farmer, who was of a reflective bent, saw here a
curious and beautiful law of nature:  ``Ten barns each with five sheep,
and then five fields each with ten sheep!"  Unfortunately this law did
not always hold, and when the wolves howled on the hill at night, it
failed quite often.  The farmer had an explanation for this:  ``The
howling of the wolves greatly upsets my sheep, and the laws of nature,
like the laws of man, are often disobeyed when agitated spirits
prevail".  The farmer realized that to make his law universal he would
have to modify it thus:  ``When tranquillity reigns, ten of five turn
into five of ten."

We today who know arithmetic would say that the farmer's law, though
true enough in his particular situation, isn't a very good law by
scientific standards.  It needs to be ``factored" into two laws, the
first being the simple and very general law that xy = yx and the second
a more complicated and specialized law having to do with sheep and
wolves.  The farmer was indeed aware that xy = yx, at least in the case
of 5 and 10, but what he could not see is that the essential condition
for xy to be yx has nothing to do with sheep or wolves or tranquillity
but is simply that the total number of sheep remain constant.  One
reason he couldn't see this is that he lacked any conception of the
total number of his sheep; that's because in those days there were no
numbers beyond ten, just ``many".

There are three morals to this tale.  The first is that it's not enough
just to ask whether a law is right or wrong---we should also ask whether
it gets to the point.  The second is more subtle:  If the point escapes
us, maybe it's because we lack the raw materials of thought needed to
even conceive of it.  The third is not subtle at all:  learn to count!

We have learned to count beyond ten sheep and even beyond three
dimensions, but we still are under a very stifling conceptual limitation
in not being able to count beyond the two types of phenomena that we
call {\it classical} and {\it quantum}.  This paper will set these two among many
more.  It will do this by teaching us some new ways to count cases, such
as how to keep counting when the count goes below zero!  This will
provide us with the raw materials for thought we need to clearly see
some crucial points that quantum philosophy has so far missed, notably
the significance, or rather the insignificance, of the wave function,
and the essentially acausal nature of quantum processes.

Quantum mechanics has revealed many puzzling patterns in nature, perhaps
the most puzzling being the EPR correlation.  The explanations we hear
for this phenomenon all too often resemble the farmer's spirit of
tranquillity.  We'll see that we can ``factor" a simple piece of
probability theory out of physics that makes sense of things like EPR in
much the same way that xy = yx makes sense of the farmer's sheep counts.
What's basically new here is that quantum phenomena in general can be
represented as simple large number phenomena whose laws belong to the
arithmetic of case counting.

But what about non-locality?  The collapse of the wave front?  Quantum
measurement?  How does the present paper fit in with the more familiar
ways of interpreting the formalism of quantum mechanics?  Let me briefly
address this question.

It was recognized quite early that quantum mechanics bears the earmarks
of a purely statistical theory.  The Schrodinger equation looks very
much like the equation governing a Markov process, and we actually get
the Schrodinger equation if we multiply the generator of a self-adjoint
continuous Markov chain by i. Now Markov processes belong entirely to
the theory of probability---there's no physics in them at all.  Could it
be that quantum mechanics does for mechanics what statistical mechanics
did for the theory of heat?  Can mechanics, and space and time along
with it, be reduced to the statistical behavior of something simpler and
more fundamental, or perhaps even to theorems in the bare science of
probability itself?

Two obstacles have stood in the way of such a simplification.

The first is that probabilities can't be imaginary.  It turns out that
we can define imaginary probabilities in terms of negative
probabilities, but that wouldn't seem to be of much help---have you ever
counted fewer than 0 cases?  But then again, come to think of it, have
you ever seen a pile of gold bricks with fewer than 0 bricks in it?  And
yet the mathematics of negative piles of gold bricks has become
indispensable for keeping accounts, especially government accounts.  So
why not ``keep accounts" with negative probabilities?  This first
obstacle doesn't look like it ought to stop us in our tracks for long,
and if it were the only obstacle, it probably wouldn't.

The second has proved more obstinate.  We saw that the Schrodinger and
Markov equations have the same form, but we must next ask, do they
govern the same quantities?  The answer appears to be no.  The numbers
that in the Markov equation are probabilities, are the square roots of
probabilities in the Schrodinger equation.  This little disparity has
for sixty years kept alive the notion of the ``wave function" whose
``amplitude" squared is probability.

And this little disparity, this seeming technicality, has been the
logjam that has kept quantum philosophy circling in the same stagnant
pool of inadequate ideas for the last sixty years.  The present paper
aims to break up that logjam.

Let's assume we have overcome obstacle 1 and can now work within an
extended theory of probability where cases count negatively as well as
positively.  As we'll see, most of the math in this extended probability
theory is the same as in the all-positive theory.  Now suppose we write
down the Schrodinger equation as a formal Markov chain, in which the
differential transition matrix operates on a state vector.  The numbers
in this vector are the amplitudes of the ``wave function", which we would
now like to think of as probabilities.  The well-known problem with this
construction is that the basic rule for quantum probability, the Born
rule, says that these amplitudes must be squared to give the
probabilities that are actually observed in quantum measurement.

However, and now we are coming to the key idea, this chain of transition
matrices we call a Markov chain is only one of many ways to represent
the joint probability distribution that is the Markov process itself.
Another is by a chain of joint probability matrices linked by
conditioning events of the form x=y.  We define the state S of such a
link as the probability matrix on x and y with the link removed.  A
state is called quantum if the distributions on x and y are identical,
and a chain is called quantum if its states are quantum and its matrices
are unitary.  Now here is the punch-line:  in a quantum chain, the
unlinked probabilities on x and y behave like quantum amplitudes, while
the linked probabilities, which are the diagonal entries of S, are the
squares of these amplitudes, and hence behave like the probabilities
they are supposed to be.  There is no wave function, only probabilities,
positive and negative.

The link method embeds quantum mechanics in the mathematics of Markov
processes in such a way that quantum amplitudes are represented by
unlinked probabilities and quantum measurement probabilities by linked
probabilities.  As we'll see, the same embedding maps classical Markov
chains into isomorphic representations of themselves.  In linked chains
of any kind, linked probabilities are quadratic in unlinked
probabilities, but in classical chains one of the factors is constant
and factors out, so probability is in effect linear in amplitude, i.e.
the classical ``state vector" is the usual probability vector.  If this
sounds a bit cryptic, don't worry, it will all be spelled out in detail.

Most discussions of the meaning of quantum mechanics these days seem to
be about the problem of the ``collapse of the wave function."  In link
theory this problem simply vanishes, since there is no wave function to
collapse.  Imagine if the Eighteenth Century caloric were still hanging
around as the official theory of heat:  we'd be chronically plagued by
ever more complicated theories explaining the collapse of the ``caloric
field" when you measure an atom's energy.  What a relief to get away
from the spell of such nonsense!

This large-number explanation of quantum mechanics raises two basic
questions:  Large numbers of what? and Must we buy it?

The answer to the first question is implicit in the above discussion,
but needs to be said simply:  The things we count large numbers of are
cases.  Simple arithmetic reveals that the core quantum laws, in a
generalized form, are features of any probabilistic system whatsoever.
Von Neumann's formulation of the Born probability rule prob(P) =
trace(PS) holds at every connection between the parts of such a system,
and the dynamical rule S'T = TS governs every part that is connected at
two places.

We brought up caloric to draw a parallel between our present situation
and the situation in physics when it was discovered that the laws
governing heat could be interpreted as statistical laws of atomic
motion.  However, there is a big difference.  In the case of heat, the
statistical theory sat on top of the Newtonian theory of motion, whereas
in our case there is no underlying empirical theory at all.  Probability
theory is just the arithmetic of case counting, so the generalized
quantum laws are like xy = yx in that their truth is assured, the only
empirical issue being where and when they apply.

The answer to the second question is no, we don't have to.  However, the
same can be said about the arithmetical explanation of five fields with
ten sheep each.  It's logically possible that when true tranquillity
reigns, the gods always make sure that every field contains ten sheep
(presumably the age of true tranquillity is long since past).  It's also
logically possible that the non-local ``guide wave" explanation of
quantum phenomena is the right one.  With both sheep and quantum, the
arithmetical explanation makes so much more sense that it would be most
malicious of the gods to reject it just to save our old habits of
thought.

We'll see that there is another reason to prefer the arithmetical
explanation, which is that, as our discussion of Markov processes
suggests, it also applies to classical things like computers.  This at
last enables us to make sense of quantum measurement, which has always
been a great mystery.  Quantum and classical now stand revealed as two
``shapes" made of the same stuff, so there is nothing more mysterious
about their both being parts of the same process than there is about
round wheels and square windows both being parts of the same car.  The
radical path also leads to a good Kantian solution of Hume's problem,
which is that of finding causality in the order of succession, and we'll
see that the choice between acausal and causal/classical thinking is to
some extent a choice of analytical method, like the choice between polar
and rectilinear coordinates.

But then comes the big question:  What about the other shapes?  The ones
other than quantum or classical that we have never before imagined, and
therefore never thought to look for in nature?  We'll briefly touch on
the big question, but it calls for a much bigger answer than we can give
here, or now.

\section{PROCESS AND SYSTEM: AN OVERVIEW}

The term `animal faith' in our subtitle is taken from the title of
Santayana's book ``Skepticism and Animal Faith"\cite{Santayana23}; it refers to what gets
us through the day and keeps our thought processes going even through
our spells of radical doubt.  Santayana was concerned to delimit animal
faith and to contrast it with other, presumably higher, things.  Our
agenda here is quite different and is closer to Kant's:  it is to
articulate and transform into explicit principles the animal faith
implicit in certain of our concepts that play a key role in science.
Once this is done, once these principles become explicit, we'll find
that they take on a new life of their own, and are full of surprises.
First, they reveal the simple mathematical structure that unifies
quantum and classical.  But then they strangely turn against themselves,
revealing their own limitations, and even rather impolitely suggesting
that perhaps we should find better ways to get through the day.  This
paper, however, is about their more cheerful messages.

Like Freud's ``talking cure", our analysis will start by paying attention
to how we talk about everyday things, and then go on to explore
subterranean labyrinths in search of hidden meanings.  But it is not a
search for what has been repressed, or for what has fallen into the
unconsciousness of habit, but for that ancient animal heritage of ``know
how" that we, as thinking people, unconsciously draw upon in formulating
our most sophisticated thoughts.

Lest this sound too ambitious, let me add that we are confining our
analysis to what underlies certain commonplace scientific words,
notably:  {\it information, part, place, event, variable, process, procedure,
system, input, output} and {\it cause}.  The unconscious beliefs we are
searching for are those that belong to the smoothly working hidden
machinery behind the easy flow of thoughts in which these words occur.
To put it another way, without certain implicitly held principles these
commonplace words, which are indispensable in any discussion of
scientific matters, would be quite meaningless, and these are the
principles we are trying to capture and articulate.

We shall see that, when these principles are precisely articulated, they
become a tightly knit system that has some surprising consequences,
including, as mentioned, the two `core laws' of quantum mechanics.  More
generally, the consequences of an analysis of animal faith are of two
kinds---call them Kantian and Freudian; the Kantian kind reinforce our
animal faith, while the Freudian kind force us to question it.

Let's begin our analysis with information.

Suppose you see something and write down what you see; this is a {\it
datum}.
{\it Data} is what has been seen and noted.  To qualify as {\it
information},
however, data must have some element of surprise.  If you know that A is
going to happen, seeing it happen may be gratifying or reassuring, but
it is not {\it informative}.  Thus {\it information} is what you have seen,
contrasted with what you {\it might} have seen.  When you {\it
describe} something,
you normally supply a number of connected pieces of information, so
{\it description} is information broken up into {\it connected parts}.

These very simple and ordinary observations are beginning to give hints
of a pattern, but they come to a halt with the difficult word `{\it
parts}'.
It's curious how in the long and contentious history of speculations
about what the world is made of, almost all the debate has been about
what are its {\it ingredients}, and almost none about the more basic problem
of what it means for these proposed ingredients to fit together as
parts.  Taking ordinary material things apart and putting them together
is such a natural activity that we blithely extend it to entities of
every kind, without ever imagining that this might lead to problems.

And this is certainly the case with information, as when we speak
confidently of ``partial information", ``the whole story", etc.  No doubt
we are at ease with such talk because the parts and wholes of
information so often coincide with the parts and wholes of the language
that conveys it.  But underlying this linguistic idea of part and whole
there is a deeper level of meaning , and excavating that deeper level
will be the main task of our analysis.

First of all, we must carefully distinguish between two kinds of part-
whole relationship:  that which we find in space and space-time, and
that which we find in material structures like buildings and computers
and molecules.  Let's call the first relationship {\it extension} and the second
{\it composition}.  The parts of an extension are its {\it regions}, and the
identity of a region, i.e. that which makes it different from other
regions, is its {\it place}.  The parts of a composition, on the other hand,
are its {\it components}, and the identity of a component is not its place but
its {\it type}.  Components are {\it interchangeable} parts; you can duplicate them,
remove them from their places and use them elsewhere, etc.  But a
region, far from being an interchangeable part, is by its nature unique;
the people of Palo Alto can leave California, even the buildings of Palo
Alto can leave California, but Palo Alto itself can't leave California
since it's a {\it place} in California.

Matter, as we currently conceive of it, is composed of elementary
particles, the most perfectly interchangeable of parts imaginable, which
makes matter a composition par excellence, which is to say, its
essential self is nowhere.  There are those today who speak of making
matter out of space or space out of matter, but this is nonsense; you
can't make extension out of composition nor composition out of
extension.  Those who claim to have a theory of {\it everything}, even if they
are right, have only found half of the Holy Grail; the other half is a
theory of {\it everywhere}!

The material objects we encounter in everyday life are always both
regions and components; they have both places and types.  Thought is a
constant dialogue between extension and composition.  Practically
speaking, the important distinction to keep in mind is between parts
that don't keep their identity when you remove them from their context,
like random variables in a joint probability distribution, and those
that do, like logic gates in a computer.  Only the latter can be used to
{\it construct} things.

In mathematics, the dialogue is between geometry and algebra.  Think of
analytic geometry, which is just such a dialogue.  In pure Euclidean
metric geometry, we start out with a homogeneous space whose places are
{\it points}, i.e. we can only identify them by {\it pointing} to them as ``here" or
``there" or ``there" etc.  In order to keep track of these points, we move
into algebra.  We do this by giving each point a unique identity as a
composition of an x-vector, a y-vector and a z-vector; this
identification scheme is known as a coordinate system.  Notice that
vectors themselves are not regions or places; you specify a vector as a
certain {\it type} of thing that can be found anywhere.  Notice also that we
must carefully specify just what it means for one vector to be a part,
i.e. a component, of another; this particular kind of part-whole
relation is what defines linear algebra as opposed to other kinds of
algebra.  Finally, notice that giving coordinates to points does not
entirely get rid of {\it here} and {\it there}; thus the dialogue continues since
we must locate the 0 vector {\it here}, and we must then point ``there",
``there", ``there" in the directions of x, y and z. Nor can we completely
get rid of {\it this kind} and {\it that kind} in geometry, since geometric
structure involves numerical ratios of distance, so we have {\it this} and
{\it that} number, and onward to {\it this} and {\it that} figure, etc.

How does the distinction between extension and composition apply to
informational structures?

Information accumulates as a progressive {\it extension} of the ``body of
knowledge".  Incoming items of information build on each other, qualify
each other, and in general, can only be understood in relationship to
their ``neighbors".  We say ``{\it this} happened, and then {\it this} happened and
then {\it this} happened .." etc., where in each case to know {\it
what} happened we
must know something about what happened previously.  From time to time,
though, we {\it abstract} an item of information from its place and give it a
kind of autonomy by turning it into a story, a design, a warning, an
example, a rule, a law, a procedure etc. that can be retold or reused in
other contexts; in short, we turn it into a component.  One way to
analyze a process is to systematically reconstruct it from components
that have been so abstracted; we'll call this a {\it reductive
analysis} and
the resulting composition a {\it system}.  Let's reflect a bit on the
distinction between process and system, which will play an important
role in interpreting our mathematical results.

The word `process', in its most general sense, means something that
`proceeds or moves along'.  But the word also has the connotation of
{\it procedure} as in {\it due process}.  Thus a process of a certain kind refers to
that which is allowed to happen, or which can happen, under certain
specified conditions.  This is how we shall construe the present
technical meaning of the word.  If the possibilities under the specified
conditions are assigned probabilities we'll call the process a
{\it stochastic} process.  More exactly, a stochastic process is a joint
probability distribution on a set of so-called stochastic variables.  In
Chapter 3, we'll focus on Markov processes, which are the simplest and
best known stochastic processes, and we'll see that the core quantum
laws as they occur in Markov processes take the form most familiar to
physicists.

A system is a process analyzed into interchangeable parts.  A good
system brings order to a complex whole by portraying it as a regular
arrangement of a small variety of such parts.  For stochastic systems
the paradigm case is a Markov chain, which is a ``chain" of connected
copies of a single part called a {\it transition matrix} T$_{ij}$.  We can think of
T$_{ij}$ as a representation of a stochastic variable j (the column
variable) whose probability
distribution p(j) is a function of a free parameter i (the row
variable). Connecting means
assigning the free parameter i to the stochastic variable j of the prior
transition matrix, thereby turning the numbers T$_{ij}$ into conditional
probabilities.  If we assign the free parameter of the first component
to an {\it unconditioned} variable, then chaining transmits the definiteness
of the first variable down the line to create a joint probability
distribution on all the succeeding variables, and the resulting process
is a Markov process.

The contrast between process and system is roughly that between
extension and composition.  A process is extended, its regions being
{\it events}.  Like many English nouns (`noun', for instance), the word
`event' can be taken either in the definite or the indefinite sense; an
event can be something in particular that happened, in which case it has
its unique place in space-time, or it can be a particular {\it kind} of
happening, as when we speak of ``the event heads" in probability theory.
One way to describe a process is as an arrangement of events in the
second sense, which we can think of as labeling its extended parts by
certain of their qualities; another word for this is {\it map-making}.  A map
is a composition of essences, or predicates, to use the modern term, but
the parts of the process that these essences identify need not be
removable components, so a map does not in general represent the process
itself as a composition.

Thus we see that there are two ways to analyze a process, or any other
extended whole:  We can {\it map} it; this is the way of the naturalist, and
also of the explorer, the historian and the astronomer.  Or we can
{\it reduce} it to a heap of autonomous parts, which we then reassemble into a
composition that has the same map the naturalist would draw.  Nowadays
we often hear about the naturalist as the good guy and the reductionist
as the bad guy.  But building things and taking things apart belong to
life as much as exploring and drawing pictures.  When the reductionist
goes bad, it's usually because the naturalist hasn't given him a good
enough map; what the reductionist then recreates from his storehouse of
interchangeable parts may resemble the naturalist's whole, but the
essence of the original is missing.

To give to a part the autonomy of a component, we must usually do more
than just copy it as it appears in place.  Rather, we must ``de-install"
it from its original context, which is to say, we must transform it into
a new entity whose features are no longer conditioned by place.  To
reductively analyze a process is to de-install its partial regions, at
least in our imagination, turning them into items that can stand alone,
and that can be duplicated in such a way that the duplicates can be
reassembled into a duplicate of the original whole.  Reductive analysis
gives the observer a more penetrating gaze, which sees not only what and
where things are but where and how they come apart.

The metaphor of de-installing and re-installing components should be a
vivid one for anyone who has had to wrestle with the problem of
de-installing an APP from Windows.  In fact, de-installing a computer
program is literally a special case of the operation we shall describe
here of {\it disconnecting} a component.  We'll go into all this in detail
later, but for now let it just serve as a reminder of how much hard work
goes into seeing our ordinary world as full of {\it things}.  This hard work
is largely unconscious, and is based on ``know-how" dating from our dim
animal past.  Our problem today is that, with the progress of scientific
thought, we have wandered into domains where this unconscious skill no
longer serves all our needs; for instance, it falls far short of telling
us how to de-install a quark from a nucleus, or an event from a time
loop, or for that matter, how to de-install a thought from the stream of
consciousness.  Thus it has become essential to dredge up and articulate
the principles we unconsciously rely on in this work of de-installing
things, and consciously learn to do it better, for otherwise we risk
disastrous encounters with things we can't imagine and therefore can't
see.

The naturalist, in contrast to the reductionist, sees things in place;
he sees what and where they are in the context of the process as a
whole.  This involves seeing extensional separations and boundaries, but
it also involves seeing how things {\it function} within the whole.  
Actually, as we'll see, functional structure is best described in terms
of the following three-place relation:

\noindent
{\bf Separability}:  We say that b {\it separates} A and C, or that A and C are
{\it separable} at b, if fixing b makes the uncertainty or indefiniteness as
to what is the case with A independent of the uncertainty or
indefiniteness as to what is the case with C.

The most familiar example of this notion is the separability of the
future from the past by the present:  If we know everything about the
present, then getting more information about the past does not give us
any more information about the future, and vice versa, or so we suppose.
(Remember, we are now examining our presuppositions, not looking for
objective truths).  Note that this says nothing about the determinism of
the future by the present---all it says is that the past and future,
however else they may be related, only ``communicate" with each other via
the present.  Which brings up another familiar example:  the
separability of two communicating parties by their line of
communication.  If A and C are talking by phone, tapping their phone
line (b) can tell you everything that's going on between them; whatever else
is going on with the two of them at the time is going on with each of
them independently.

Another term for separability is {\it conditional independence}.  In the
theory of\break 
stochastic processes the two terms have essentially the same
meaning, though we'll keep them both to refer to slightly different
mathematical formulations (see Chapter 2.)

\noindent
{\bf Independence, unconditional and conditional}:  Events A and C are called
{\it independent} if the probability of A\&C is the probability of A times the
probability of C. A and C are called {\it conditionally independent} given
condition B if they are independent in the probability distribution
conditioned by B, i.e. if p(A\&C$\vert$B) = p(A$\vert$B)p(C$\vert$B).  If for some random
variable x, A and C are independent given condition x=k, where k is any
value of x, then we say that x {\it separates} A and C. In the above telephone
example, if we regard b as a random variable, then b separates A and C.

Since every probability distribution is conditioned by something, there
is a sense in which all independence is conditional.  The relation of
independence, like every notion of being different or separate, is
really a three-term relation:  A and C are independent in a certain
context B. We have just considered the case where that context is the
result of placing a certain condition on a probability distribution.
But the more familiar case is that in which the context is the result of
{\it removing} a certain condition on a probability distribution.  Notice that
although there is only one way to impose a condition, there are many
ways to remove a condition; this is one reason why there are so many
different ways to represent a given process as a system.  The converse
of ``deconditioned" independence is:

\noindent
{\bf Conditional dependence}:  We say that A and C are {\it conditionally
dependent} given B if p(A\&C$\vert$B) is unequal to p(A$\vert$B)p(C$\vert$B).

A very important kind of conditional dependence is that in which the
condition B is of the form x=y, where, without B, A\&x is independent of
y\&C.  This is called linking, and we'll return to it shortly.  There is
a fundamental theorem that relates
linking to separability:

\noindent
{\bf Disconnection theorem}:  Process A\&x\&C is separable at x if and only if
there exists a process in which A\&x is independent of y\&C which reduces
to A\&x\&C under the condition x=y. This, and other theorems stated in
this chapter, are proved in Chapter 3.

Speaking more generally, what makes separability so important is that it
always coincides with the possibility of breaking a process into two
de-installed components of some kind.  We separate A and C by listening
in on their phone line b, which is at the same time their connection and
their common functional boundary, at least as far as their conversation
goes.  On the other hand we {\it disconnect} A and C, turning them into
autonomous components, by {\it cutting} their phone line; this ends their
conversation because their phone line is their separation boundary.
When we mark the separation boundaries in a process, we may be
functioning as naturalists, but we are also like the butcher with his
blue pen marking the chops in a carcass.

To reconnect A and C so as to restore the whole process is to equate
what is the case with b1 and what is the case with b2, thus restoring
the original situation of A\&b\&C.  Just what it means to equate b1 with
b2 turns out to be less obvious than it sounds, and in fact can be
understood in two very different ways; we'll call these two ways {\it i-o
connection} and {\it linking}.  Understanding exactly how an i-o connection
differs from a link is the key to understanding how quantum processes
differ from classical processes such as phone conversations.  Let's
start with i-o connection.

We'll simplify slightly by supposing that A is talking to C on a one-way
line b. If that line is cut, what happens to the signal voltage at the
two cut ends b1 and b2?

The cut line b1 from A is an {\it output}.  This means, ideally, that what
comes down the wire depends only on A, so the signal at b1 is unchanged
by the cut.  The cut line b2 from C is an {\it input}, however, and our
description of the situation doesn't specify how inputs behave when they
are disconnected.  An open line will normally produce white noise,
sometimes mixed with hum and faint radio signals.  However, and this is
a crucial point, the designer of a system usually doesn't have to take
into account disconnected inputs; he only needs to think about how the
system behaves when inputs are ``driven" by outputs, whether from the
user or from other components.  To put it another way, the designer can
regard inputs as {\it free variables}; his components are {\it open
processes},
analogous to open formulae in the predicate calculus.

Open processes are interconnected by equating inputs to outputs, which
means {\it assigning} free variables to non-free or {\it bound} variables (assigning
variables is a familiar concept for computer programmers.)  A
composition created in this way will be called an {\it i-o system}.  If all
the free variables of the components are assigned, we call the system
{\it closed}; otherwise it's an {\it open} system.

We are on familiar ground here; this is how engineers and programmers
are taught to think these days.  Once again, here are the main ideas:

\noindent
{\bf Input}: A free variable in an open process.

\noindent
{\bf I-o connection}:  The assignment of a free variable to a bound variable.

\noindent
{\bf Output}:  The bound variable to which a free variable is assigned.

Disconnecting is just the reverse of connecting.  More exactly:

\noindent
{\bf I-o disconnection}:  A and C are disconnected at variable b means we are
given two processes A\&b1 and b2\&C with b2 free such that when b2 is
assigned to b1, the result is the original process A\&b\&C.

\noindent
{\bf I-o system}:  A composition of connected open processes.  Every i-o
system represents a process, but we must be careful to distinguish this
process from the system, just as we must be careful to distinguish a
vector from its representation as a sum in a particular basis.

There's something not quite right about all this.  After all, a free
variable is only an abstract linguistic entity, whereas our cut wire,
lying there on the ground and picking up white noise, remains very much
a real material object.  The parts of a telephone system are not
``indefinite processes"---they are just as definite as houses and trees
and stones.  Or so says animal faith.  There is in fact a way to
describe the situation that's more in keeping with this bit of animal
faith, which is to replace the concept of assigning a free variable by
the concept of {\it linking} a so-called {\it white} variable.

\noindent
{\bf White variable}:  A random variable on which there is a uniform
probability distribution.  In the case of a discrete variable, whiteness
means that every value is equally probable, and hence the variable has a
finite number of values, since their probabilities must add up to 1. A
white real variable has a uniform probability density which must
integrate to 1, so it's range is bounded.

\noindent
{\bf Theorem}:  If x and y are independent random variables with the same
range, and y is white, then the condition x=y on their joint probability
distribution does not change the distribution on x.

In our telephone example, let's now forget about free variables and
think of the components A\&b1 and b2\&C as closed processes.  Taken
together, the two can be regarded as independent regions in the process
A\&b1\&b2\&C.  When we connect b1 to b2, the result is a new process A\&b\&C.
How does this new process differ from the old?  The obvious answer is
that it is just like the old one except for its behavior being
restricted by the condition that b1 equals b2.  Connecting two wire ends
means making their signals equal.

It is not immediately clear how this answer, which is couched in the
physical language of broken wires, signal voltages etc., applies to
processes as such.  (Remember, a process, as we are now defining it, is
simply a joint probability distribution on a set of variables.).
However, the phrase ``.. restricted by the condition.." gives a strong
hint.  To say ``the probability distribution D1 is just like the
probability distribution D2 restricted by the condition C" means that
you get D2 from D1 by conditioning all event probabilities in D1 by
event C. Thus our answer is basically this:  To connect any two
variables x and y, condition all of the probabilities of the
disconnected system by the event x=y.  This particular bit of animal
faith brought to light will turn out to be the key to making sense out
of quantum mechanics, so let's forthwith coin words that nail it down:

\noindent
{\bf Links}:  Place a condition of the form x=y on two variables x and y of a
stochastic process, thereby creating a new process in which the
unconditional probability p(E) of any event E is p(E $\vert$ x=y) in the old
process.  This condition is called a link.

\noindent
{\bf Link system}:  A stochastic process plus a set of links.

We'll go over all of this in considerably more detail in chapters 2 and
3. A summary of the relevant concepts, terminology and notation of
probability theory will be found in Chapter 3, Section 1. Let's now
briefly look at some high spots.

First, if x and y are independent and y is white, then the link between
them behaves exactly like an i-o connection, with y acting like the
input; this follows immediately from the above theorem.  We'll show in
Chapter 3 that we can neatly translate any i-o system into a link system
which represents the same process and has essentially the same formal
description.  Engineers should have no problem adapting to link
descriptions.

Second, and this is extremely important, we'll prove that this natural
mapping of i-o connections to links does not go backward!  Though every
i-o system is equivalent to a link system, most link systems have no i-o
equivalents, and even when one can make an i-o model of a link process,
that model is usually exponentially more complicated than some link
representation of the same process.  Quantum systems are among those
having no i-o counterparts, though quantum measurement systems do, as
we'll see in Section 3.7. If technology can learn to really
deal with this larger class of non i-o systems, it will turn into a
fundamentally new kind of enterprise.

Third, a crucial definition, that of a {\it link state}:

\noindent
{\bf Link state}:  {\it Given} any link x=y, the {\it state} of that link is defined as
the matrix representing the joint probability distribution on x and y
without the link (actually, it's this matrix normalized by dividing by
its trace, but that's a small detail.)  This is the concept that
replaces the quantum wave function, and does away with the problem of the
collapse of the wave function and other related quantum nuisances,
as we discuss later.

In the introduction we briefly mentioned caloric.  Heat, in the
Eighteenth Century, was regarded as a fluid called caloric which was
somehow related to the mechanical behavior of objects, but subject to
its own non-mechanical laws.  A series of experiments during the first
half of the nineteenth century, starting with Count Rumford's crude
qualitative observations, and ending with the definitive work of Joule
and Helmholz, convinced physicists that they didn't actually need to
postulate a new substance called caloric, since heat is simply energy.
When it was further realized that this energy is the random kinetic
energy of molecular motion, thermodynamics became statistical mechanics.

History to some extent repeated itself in the early twentieth century
with de Broglie's wave mechanics.  His ``matter waves" were at first
thought to be waves in some new kind of fluid, the quantum analogue of
caloric, but after Born discovered his quantum probability rule, they
lost most of their substantiality, and physicists began to call them
``probability waves".

Quantum mechanics, however, didn't go the way of thermodynamics.
Because the so-called probability wave is really a wave of the square
root of probability, ``quantum caloric" didn't seem to be reducible to
anything already known.  And, alas, it is still with us, ignored by most
working physicists, but hanging around the edge of physics in a kind of
limbo where from time to time it is reworked and touted as a wonderful
new discovery.  But fortunately this unhappy state of affairs is almost
over.

Here, in a nutshell, is how link states reduce ``quantum caloric" to
something already known, namely probability:

Suppose x and y are independent random variables.  Then the probability
that x and y will both have some value k is the product of the
probabilities that they will have k separately, i.e. p(x=k \& y=k) =
p(x=k)p(y=k).  Let's call p(x=k) and p(y=k) the unlinked probabilities
of x being k and y being k. Now suppose we impose the link condition
x=y.  The linked probability of x=k is then the probability, conditioned
by x=y, that both x=k and y=k.  The probability that x is k, as a
function of k, is proportional to p(x=k)p(y=k).

In short, linked probabilities are always quadratic in unlinked
probabilities.  If the distributions on x and y are identical, which is
the quantum situation, then linked probabilities are the squares of
unlinked probabilities.  That's essentially all there is to it!  Quantum
amplitudes are not the intensities of some mysterious new fluid, but are
simply unlinked probabilities.  One might ask ``Probabilities of what?",
but we'll see that we don't have to answer this question, just as we
don't have to answer the question of what sets of things have numbers x
and y in order to understand the meaning and truth of xy = yx.

A generalized form of quantum amplitudes will be found in any
statistical situation whatsoever if we subject that situation to link
analysis.  Quantum amplitudes proper will show up if probabilities can
go negative (we'll return to this in a minute) and we de-install
components in such a way that the link states possess quantum
symmetries; as we'll see in Chapter 3, this is always possible, though
not always advisable.

Link states are quantum states represented as von Neumann density
matrices.  However, we can start in the more familiar way with quantum
states as vectors over the amplitude field, and when we apply link
analysis to the Schrodinger equation, amplitudes still turn out to be
unlinked probabilities.  Here, very roughly, is why this is so; the
details are in Chapter 3.

In a continuous Markov chain, regarded as an i-o system, the probability
distribution on the state variable can be represented as a vector which
evolves in time according to the law v' = T(v), where v is the
distribution vector at time t, v' the vector at t+dt, and T the
differential transition matrix.  The Schrodinger equation, expressed in
the language of Hilbert space, has exactly the same form, the difference
being that the components of v are amplitudes rather than probabilities,
and T is a unitary matrix rather than a transition matrix.

Let's now represent the Markov chain as a link system rather than an i-o
system.  The changing state vector is now the diagonal of a link matrix
S that changes according to the law S' = TST$^{-1}$. If we think of S and S'
as density operators, this is the von Neumann generalization of v' =
T(v).  An immediate corollary is the Born probability rule in the
generalized von Neumann form prob(P) = trace(PS).  The rule for turning
an i-o system into a link system gives S the form $\vert$ v $><$ w $\vert$, where w is a
``white" vector, i.e. it represents a uniform probability distribution,
so it is only v that varies with time.  The diagonal entries are
quadratic in the entries of v and w, but since the constant entries of w
are equal they factor out, so the probability vector is just v, the same
probability vector that occurs in the i-o system.

Next, let's do the same with the quantum chain.  Again we have S' =
TST$^{-1}$, but now T is unitary.  And again we have S = $\vert$v$><$w$\vert$, but now
instead of w being white, we have w=v, which makes the diagonal entries
into the squares of the entries in v, i.e. state probabilities are the
squares of  ``amplitudes"!  The amplitudes are no longer the values of
some mysterious wave function but are simply the probabilities in the
joint distribution on a pair of unlinked variables.  Quantum caloric is
nowhere in sight!

That takes care of the main obstacle to a statistical interpretation of
quantum mechanics.  The square law is no longer a reason not to regard
probability waves as true waves of probability.  It's true, the wave
amplitudes aren't the probabilities that we measure; rather, they are
the probabilities that we would measure if the quantum process were
disconnected at a separation point.  But the same is true in a Markov
chain, analyzed as a link process, the only difference being that in the
Markov chain the output probabilities don't change when the output is
linked to an input.  In a quantum chain, ``input" and ``output" have
identical distributions, and since unlinked state variables are
independent, linking yields the unlinked probabilities squared.

There is still the problem of amplitudes going negative (they can also
go imaginary, but as Mackey observed, the scalar i can be interpreted as
a unitary symmetry in real quantum mechanics)\cite{Mackey63}.  What
could possibly be the meaning of negative probabilities?  How can there
be fewer than 0 things having a certain property P?  How can an event E
occur less than 0 times?

These are not questions that can be answered by constructing some
ingenious mechanism, since they involve logic itself.  Earlier we
compared negative probabilities to negative piles of gold, my point
being that there is no problem in constructing a useful mathematics that
incorporates them.  But that doesn't tell us what they really are, and
we
now doubt very much that their deeper significance can be understood at
all within the scope of science as we know it.

This could be enough to drive many people back to quantum caloric.  But
before you join the stampede, be aware that the logical mysteries in
quantum mechanics don't come from the link state interpretation.  They
come from the bare fact of amplitude cancellation, and practically stare
you in the face with EPR.  How can it be that A is possible, and also
that B is possible, but that (A OR B) is impossible? ---that's the
mystery of the two-slit experiment.  
Here ``OR'' is the logician's ``or'' which allows A
and B both to be true, A true and B false, A false and B true,
but requires (A OR B) to be false if both are false.
The double slit example we have is mind is: 
Case A: slit A open leading to a detection; Case B: slit
B open leading to a detection; case (A OR B): 
both slits open but there is no detection. Von Neumann was well aware of the
logical strangeness of quantum phenomena, and tried to cope with it by
means of a non-Boolean construction that he called quantum logic.  This
turned out to be a dead end, but other strange logics dating from the
period of quantum logic show more promise.  Enough of this for now,
however.  We'll return briefly to the logic of negative case-counting in
Chapter 2, but it's too big a subject to do justice to here; suffice to
say that the logical anomalies don't vitiate our simple mathematics.

So where do we stand with quantum mechanics?  We've gotten rid of
``matter waves" but the mystery of quantum interference remains, since to
derive the Schrodinger equation in link theory does require negative
probabilities, even in the process itself.  But now, at least, the
mystery of interference has become a {\it simple} mystery, and there is no
longer any reason to think of it as belonging to physics in particular.
Once one has really grasped how universal the two generalized quantum
laws are at the case-counting level, it's as hard to take seriously the
arcane ``models" of quantum mechanics that abound today as it is to take
seriously the ancient sheep farmer's ``spirit of tranquillity" after one
has understood that xy=yx.

But we're afraid we are getting somewhat ahead of our story.  We started out
to psychoanalyze animal faith, to dredge up some of the unconscious
beliefs that underlie our scientific thinking.  ``Wait a minute!", you
protest, ``I don't know about you, but I can't find any of this fancy
mathematics in my unconscious!"  Fair enough, neither of the authors
can either.  It's the
analyst's job to propose, not to pronounce, and his proposals must
always be confirmed by the analysand, who in this case is all of us.
However, what we're looking for in our unconscious is certainly not
fancy mathematics.  Animal faith in itself, i.e. in place, can't even be
put into words.  To ``find" it means to articulate something that does
the same job, namely that of supporting our habitual modes of conscious
thinking.  In our current jargon, to articulate an item of animal faith
means to de-install it, i.e. to transform it into a proposition that can
stand alone.  However, once articulated and understood, this new
creation can then be put to the test by re-installing it in our
intuition:  Does it ring a bell?  Do we think ``Ah, yes, that's what I've
always thought, even if perhaps I never quite said it that way."?  If
the items we have articulated pass this kind of test, then we can
confidently move on to the interesting surprises.

Quantum mechanics is one such interesting surprise; another, which is
the topic of Chapter 2, is a very Kantian answer to Hume's skepticism
about the existence of causes.  Kant's own answer was that we
necessarily see causes everywhere because, to paraphrase the way
Bertrand Russell put it,
we wear ``causal-colored glasses" \cite{Russell45}.  By combining the
philosopher von Wright's insight into how we operationally define
causality with our analysis of process and system, we'll bring this
statement up to date:  We necessarily see input-output systems
everywhere.  Since input-output systems are closely related to computer
models, it could be said that we necessarily see computers everywhere -
we all wear computer-colored glasses!

If all this sounds like the grand march of progress, be aware that the
two surprises above are not entirely harmonious.  What makes our Kantian
analysis of causality possible is the fact that the separability
structure of a process only partly determines the form of the link
matrix S, and we can show that it is always possible to choose a causal
S (i.e., an S with an input).  However, quantum S's are not causal!
Furthermore, and this is crucial, the choice of S for a given
disconnection will, in general, affect the separability structure of the
disconnected parts.  Thus things that come apart neatly with one type of
S may exhibit bad ``non-localities" for another type of S. EPR comes
apart very neatly with a quantum analysis, but any causal analysis
compatible with the data will exhibit non-locality, as Bell's theorem
shows.  Thus, though we are always perfectly free to choose causal S's
and thus turn the universe into a giant computer, this computer may have
a gosh-awful tangle of long distance wiring that could be eliminated by
using other S forms.

Kant has had his say---now it's Freud's turn.  Freud's project was not
to justify our beliefs but to change our minds.  He was, of course,
working with people whose minds were very much in need of changing
(which of course doesn't include you and me).  Still, he may be able to
give us a few useful tips.

As long as our assumptions remain unconscious, we have no choice but to
believe them, or rather, we have no choice but to act as if we believed
them.  The question of belief doesn't actually arise until we are able
to contemplate and weigh alternatives.  For this to happen with some
article of animal faith, we must first of all raise it up from the
unconscious depths in such a way that it becomes a proposition, i.e. we
must de-install it.  Then one of two things may happen.  We may take a
clear look at this proposition and recognize its absurdity:  ``Wait a
minute, my mother isn't really threatening to lock me in my crib if I
try to go outdoors!", or whatnot.  Or, we may see that the proposition
is obviously true.  If this happens, then another question arises:  is
this truth ``objective", or does it depend on how you look at it?

How do we answer this last question?  How do we find out whether
something depends on how we look at it?  One way is to see whether we
can look at it differently.  If we believe that all print is too fuzzy
to read, we might try on glasses.  Of course we may have to experiment
to find the right glasses.  Kant's error was to confuse our glasses with
our eyes, so-to-speak.

The Kantian conclusion of Chapters 2 and 3 is that we all wear causal
glasses, which turn out to be the same as computer-colored glasses.
Does this mean that we are forever fated to see everything as a
computer, or can we take these glasses off?  As mentioned, ``causal
glasses" correspond to a particular kind of link matrix.  We can work
mathematically with very different kinds of link matrices, and indeed we
must do so in order to practice quantum physics.  Chapter 3 will make
this very clear, and will point us in the direction of an expanded kind
of science that works with the full range of acausal state matrices.
But the question still remains:  Can we, as human beings living our
daily lives, ever take off our causal glasses?

For mathematicians, all things are possible.  The experimenter who is
willing to push hard enough may occasionally force nature to reveal its
unnatural proclivities.  But you and I, when we are just trying to get
on with it, must constantly ask questions whose answers begin with
``Because". That's our human nature.  But what will be the effect on how
we live and act if we can learn from the coming science how to go beyond
``Why" and ``Because"?  What sort of beings might we someday become?

\section{HUME'S PROBLEM}

All modern discussions of causality begin with Hume.  It was Hume who
first clearly pointed out that merely knowing the regular succession of
events does not tell us what causes what.  The fact that B always
follows A gives us no grounds for concluding that A ``forces" B to
happen.  Why speak of causes at all, then?  One school of thought says
we should stop doing so; here is the young Bertrand Russell on the
subject\cite{Russell12}:

``... the reason why physics has ceased to look for causes is that, in
fact, there are no such things.  The law of causality, I believe, like
much that passes muster among philosophers, is a relic of a bygone age,
surviving, like the monarchy, only because it is erroneously supposed to
do no harm."  

The mature Bertrand Russell, however, found out that like the rest of us
he was unable to say very much outside of logic and pure mathematics
without bringing in causes.  Cause and effect pervades everyday life;
the solution to Hume's problem can't be confined to the ivory tower of
philosophy.  A recent news item reported the discovery of a correlation
in Mexico City between heavy chili pepper consumption and stomach
cancer.  For us chili pepper lovers, this was most unwelcome news:  Do
chilies cause cancer, or might there be some other explanation?  What's
at stake here is no mere philosophical abstraction; it's whether we have
to stop eating chilies!

If there is no causal link between A and B, then, even if there is a
high correlation between A and B, our wanting to bring about or prevent
B is no reason to do anything about A. But if we learn that A causes B,
or that it significantly increases the probability of B, then whatever
power we may have over A becomes power over B too.  The understanding of
cause and effect is essential for practical life, indeed it is essential
for our very survival, because it is the knowledge of means and ends.

This obvious fact of life has been curiously neglected by philosophers.
The logician G. H. von Wright, however, was an exception; he actually
went so far as to propose that the ends-means relation be used as the
defining criterion of causality \cite{vonWright74}.  
That is, he would have us say that, by definition, A causes B if
and only if A can be used as a means to B, and more generally, that
there is a causal connection from A to B if and only if we can change B
by changing A. Notice that this is an operational definition:  it says
that the presence or absence of a causal connection between A and B can
be detected by performing a certain operation, which is to change A and
notice whether there is a concomitant change in B. That is, of course,
just how diet scientists would go about seeing whether chilies cause
cancer.

Does von Wright's definition solve Hume's problem?  It certainly gives
causality an empirical anchor that cannot be neglected in any proposed
solution.  However, a very important question still remains:  Is it
possible to find formal patterns in the data itself that correspond to
the causal connections revealed by the dispositional relation of
means-to-ends?  This is a difficult and subtle question, and the search
for its answer ultimately takes us into territory where the question
itself essentially disappears.  First, though, there is a lot of work to
be done closer to home.

The people who really live with causality are not philosophers but
engineers, and curiously enough, they rarely use the word ``cause".
That's not because, as Russell said, ``there is no such thing", but
because they have developed a much richer and more discriminating
vocabulary for the various causal structures and relationships they deal
with.  `Variable, `input', `output', `process', and `function' are
standard causal terms.  Computer science has added some important
newcomers such as, `assignment', `call', and `subroutine'.  Indeed, the
concept of subroutine neatly revives the whole Aristotelian foursome of
causes:  The material cause of a subroutine is the computer it runs in,
the formal cause is its source code, the efficient cause is its call,
and the final cause is the purpose it serves in the calling program.

It's clear from this brief survey of terms that, for the engineer,
causality is an aspect of how something is composed, not of how it is
extended.  The engineer's job is to put together certain components into
a system that realizes a causal process, i.e. a process having inputs
that produce certain desired outputs.  Furthermore, the components
themselves are causal processes, and putting components together means
using the outputs of certain components to causally control the inputs
of others.  As remarked in Chapter 1, the information engineer does not normally
concern himself with the behavior of unconnected inputs. For him, and
more generally for low power transfer, the
behavior of input B connected to output A is usually the behavior of A in
the absence of the connection, so B can be regarded as an indeterminate
or {\it free} parameter.  Thus, what the engineer designs, in our current
jargon, is a functional i-o system; its components are what we called
{\it open processes}.  The functional dependence of outputs on inputs is
called a {\it transfer function}.  The concept of a transfer function clearly
matches von Wright's concept of causality as the potential for the
relationship of means to ends, so it makes sense to refer to functional
i-o systems as {\it causal systems}.

In a causal system, so defined, the outputs of each component are
functions of its inputs, and their composition by assignment of inputs
to outputs is functional composition.  However, we now know that, in the
case of macroscopic components, the functional dependence of output on
input is a large-number effect resting on the statistical behavior of
atoms.  A more accurate, and more general, account of the engineer's
systems makes the output of a component into a {\it probabilistic} function of
the input, with the causal transfer function replaced by a transition
matrix of conditional probabilities; these are the i-o systems of
Chapter 1. The composition of transfer ``functions" is no longer
functional composition but matrix multiplication, which reduces to
functional composition in the special cases where the transition
matrices contain only 0's and 1's.  Having carefully noted the
distinction between the mathematician's function and the engineer's
transfer function, we'll now relax our language a bit and use the word s
``cause" and ``causal" in connection with both.  Thus we'll often refer to
i-o systems as causal systems and to i-o states as causal states,
especially in a context where we are distinguishing these from quantum
systems and states, which are essentially a-causal.  A more accurate
qualifier would be 'statistico-causal', but it's too much of a mouthful.
Since ``classical physics'' is often taken to imply strict determinism,
we emphasize that the ``classical'' systems we consider in this paper
{\it always} contain stochastic elements.

Suppose an input is connected to an output---how do you know which is
which?  To find out you disconnect them; the output is the one that
continues to do by itself what they did together when connected -
another operational test of causality.  But suppose you can't disconnect
them, and suppose also that you can't manipulate the inputs to the
system; all you can do is watch.  Then how do you identify causality?
Is there nothing we can learn about what causes what by just watching?
Hume's problem is still with us.

Here's another take on Hume's problem:  Are there patterns in the joint
probability distribution of an observed process that mark the causal
relations we would see if we were to experiment with its disconnected
parts?  To put it another way, do cause and effect belong to the process
itself, or are they artifacts that result from taking the process apart?

The theory of link systems throws new light on this question.  Recall
the disconnection theorem, which says that a process can be disconnected
or ``de-linked" at x if and only if it is separable at x. This shows us a
way to transform a process into a link system:  First disconnect x into
x1 and x2, where x2 is ``white" (corresponding to an input), which
produces two independent processes, call them A and B. Then link A and B
by imposing the condition x1=x2; this gives us a two-component link
system exhibiting our original process.  Now do the same for separating
variables in A and B etc. until there are no separation points left.
The result is a system which ``factors" the original process into ``prime"
components that cannot be further subdivided.  Recall that separability
belongs to the process itself, i.e. it is intrinsic to the probability
distribution . Thus the disconnection theorem reveals severe limitations
on how we can construe the correlations in a process as causal.  A
highly correlated joint probability distribution need not have any
separating variables at all, so causality requires something besides
mere correlation in the data itself.  We have at least a partial answer
to Hume.

We'll see in Chapter 3 that the earmarks of causality in the data itself
can actually be quite distinctive.  Given the temporal order of the
events in the process, it turns out that there is a unique prime
factorization into components with white inputs.  We learned in Chapter
1, and it will be proved in Chapter 3, that such systems are in natural
1-1 correspondence with i-o systems representing the same processes.
Thus from a knowledge of the given alone, i.e. of the statistical
behavior of the system without our interference, we can completely
describe the i-o components we would get if we took the process apart at
its (intrinsic) separation boundaries.  The cause and effect relation,
which we first located in the transfer functions and interconnections of
the parts, is already {\it there} in the process itself!

OK, Mr. Hume, what do you say to that?

``Not so fast, sir; you were too glib with your prime factorization.  Why
should the x2's be white?  And, if you regard what is given to our
understanding to be only the probability of certain events, then on what
basis do you bring in temporal order?  If you must call upon such
invisible help to reveal causality in the bare order of correlation, you
do but reinforce my conviction that causes are phantoms of the
imagination."

Good points.  Let's translate them into the modern idiom.  It will help
to focus our ideas if we confine ourselves to processes in which the
separating boundaries are all in a line; these are called Markov
processes.  Their definition can be stated in terms of a three-term
relation among events called the Markov property, which is defined as
follows:

\noindent
{\bf Markov property}:  To say that the triple of events (A,B,C) have the
Markov property means that they satisfy the equation 
p(C $\vert$ A\&B) = p(C $\vert$ B),
which we'll refer to as the {\it Markov equation}.

\noindent
{\bf Markov process}:  A process in which the variables are indexed by a time
parameter such that, at any time, if A is any event not involving future
variables, and B is the state of the present, and C is any event not
involving past variables, then A, B and C satisfy the Markov equation.

Recall that the probability p(X $\vert$ Y) of X conditioned by Y is defined as
p(X\&Y)/p(Y).  Bearing this in mind, notice that if we multiply both
sides of the Markov equation by p(A\&B)/p(B) we get the equation p(A\&C
$\vert$ B)
= p(A $\vert$ B)p(C $\vert$ B).  This relationship among A, B and C is known as
{\it conditional independence}; in words, A is independent\break 
of C, given B.
Recall that in Chapter 1 we referred to separability as conditional
independence; this equation defines what we were referring to.  Event B
separates events A and C means that, given B, the probability of A\&C is
the product of the probabilities of A and C. It turns out to be more
convenient to define separability by a slightly different equation,
which is gotten by multiplying both sides of the Markov equation by
p(A\&B)p(B).

\noindent
{\bf Separability}:  To say that B separates A and C means that 
p(A\&B\&C)p(B) = p(A\&B)p(B\&C).

If none of the joint probabilities of A, B and C are zero, then the
Markov property, conditional independence, and separability are all
logically equivalent.  However, for conditional independence to make
sense, p(B) cannot be zero, while for the Markov equation to make sense
p(A\&B) must also not be zero.  Since separability always makes sense,
whatever the probabilities, it's the most convenient of the three to
work with.  Such fine points needn't concern us now, however; the
really important point is this:

Conditional independence and separability are symmetrical in A and C,
which means that the defining property of a Markov process is
symmetrical in time!  Think about that for a minute.

The Markov equation seems to have a temporal arrow, since it's about
conditioning of the future by the present and past, but we now see that
this is an illusion.  Our argument for causality being in the process
itself rested on a theorem that assumes a given temporal order of
events, and in the usual discussions of Markov processes, temporal order
is taken for granted.  But Hume's challenge was to find causality in the
bare order of correlation.  We now see that this bare order does not
distinguish past from future, so it would seem that Mr. Hume is right:
Cause and effect really are in the eye of the beholder.

Kant, who has been listening to this discussion with growing impatience,
can restrain himself no longer.  Since his style is notoriously obscure,
we will freely translate his remarks, with a little help from the eminent
historian of philosophy Harald Hoffding\cite{Hoffding55}.

``We gather that by `process' you mean something that is accessible to
experience.  Now experience, in contrast to mere imagination, is a
complex composed of elements some of which are due to the faculty of
knowledge itself, while others are the result of the way in which this
faculty is determined to activity from without.  As I have shown, space
and time are forms of our perception.  For whatever the nature of our
sensations, and however much they may change, the spatial and temporal
relations in which their content is presented to us remain the same; a
space or a time does not change, whatever be its filling out.  It
follows that the temporal order of the process is given, since it
belongs to the form whereby the outer determinant of our experience of
the process can manifest itself in experience as a process.  Hume is
correct that cause and effect are in the `eye of the beholder', but
wrong to imagine that we could see otherwise than as we do."

Kant's ambition was to describe once and for all the ``forms" that belong
to the ``faculty of knowledge", leaving the ``outer determinants" of
experience in a limbo where they have become known as ``things in
themselves".  Today we are more modest.  For one thing, we no longer
make his absolute distinction between the subjective and the objective;
rather we see objectivity as a matter of degree.  Also, instead of
trying to directly analyze the faculty of knowledge, we often rest
content with characterizing the contribution of the subject to an
experience in terms of a ``choice of viewpoint", and of the object as a
particular feature of the experience that is invariant under the choice
of viewpoint.  The subjective-objective boundary is of course quite
mobile, and depends on the range of possible viewpoints we are
considering.

Even transposed into a modern setting, Kant's observations about time
direction are telling.  So far we have been taking the notion of process
as simply given.  But we must ask:  given to what?  If given merely to
imagination, then of course processes can be anything we want them to
be, and they can go backwards or forwards or any which way in time.  But
if we are empirical scientists, then processes are something given to
{\it experience}, at least potentially, which means that in thinking about
processes going backwards we must consider whether we could actually
observe such a process.  And here we must heartily agree with Kant that
the answer is no, since experience itself has an inexorable temporal
arrow, whatever its content.  Indeed the process as something given to
experience is given one piece at a time, and the order in which these
pieces are given is precisely the temporal order that we find in the
process itself.  This being the case, what could it possibly mean that
time goes backwards?

We've been describing the process itself as what is given, but here's
something else to keep in mind:  The time reversal of the {\it given} is the
{\it taken away}!

Let's be more concrete.  Suppose we are observing a process and writing
down what we observe.  Our observations accumulate in a diary which
starts with 100 blank pages and progressively fills with writing.  Now
each letter that we write reduces the nominal possibilities for what the
diary {\it could} contain by a factor of 26; this is the Shannonian sense in
which it constitutes information.  Thus the process itself proceeds
step-by-step alongside of a process of information gathering, which
progressively reduces the vast initial range of possible states of the
diary to one state, the chronicle of {\it what actually happened}.

We decided in Chapter 1 that it is not a particular chronology that we
will call a process, but rather the conditions and rules that govern it.
Let us then imagine an enormous series of repetitions governed by the
same conditions and rules, each producing a diary of what actually
happened.  We now are given an enormous heap of diaries from whose
statistics we abstract the joint probability distribution that we having
been calling a stochastic process.  The question is whether the
stochastic process so abstracted exists as a thing-in itself apart from
its potentiality of ``determining our faculty of knowledge to activity",
namely, the activity that produces this heap of diaries.

The Kantian answer is no.  The very {\it idea} of process is inseparable from
the idea of a progressive accumulation of observed information in a
diary.  Such a progression has of course a time arrow, that of the
progressive narrowing of a range of possibilities.  Thus a Markov
process backwards in time is nonsense.  To be given a process without a
time arrow is inconceivable, since the given-ness of a process is
conceptually tied to the given-ness of successive stages of its
presentation to a potential observer.

It took a lot of fancy footwork, but time and with it causality seem at
last to have returned to their familiar shapes.  Finally we are back
safe in the bosom of animal faith.  Or are we?

Kant says it's OK to believe what our animal faith tells us, because we
can't help it ---such beliefs belong to the very ``faculty of knowledge".
Darwin's theory of evolution also suggests that it's OK, but for a
somewhat different reason, namely that our faculty of knowledge evolved
to enable us to survive in an all- too-real world of hostile and
dangerous ``things-in-themselves".  The two messages, despite their
opposite philosophical starting points, are actually not all that far
apart.  Darwin's message applies an important corrective to Kant's,
however, which is that in analyzing our faculty of knowledge we must
take into account the special conditions that obtained in the
environment where that faculty evolved.  Thus, the belief that the world
is flat is certainly ``true enough" for creatures who never travel more
than a few miles from where they are born, but is badly wrong for
long-distance navigators.

Homo Sapiens is perhaps alone among the animals in being able to modify
his animal beliefs in response to drastic changes in the environment,
including those of his own making.  Furthermore, and in this we humans
are probably unique, our faculty of knowledge includes a faculty of
abstract reasoning that can extrapolate to conditions far beyond our
present environment.  Sometimes this precedes a big move:  think how we
used every branch of modern science to prepare ourselves for exploring
and colonizing outer space.

Kant lived in the Age of Enlightenment whose idols were Euclid and
Newton, and rather naively incorporated their systems into his
``intrinsic" faculty of knowledge, thus confusing a brief phase of human
culture with something final and absolute.  One wonders what he would
have said about Einstein's theory of relativity.  Kant thought it was
impossible to experience space as non-Euclidean, and yet today we have
good reason to believe that certain regions of space are quite
non-Euclidean, and we know at least some of the ways in which this
impinges on experience.  The most drastic departures from Euclid involve
warps not only in space but in time, producing loops where time closes
back on itself.  For the inhabitants of such a loop, if there be such,
the future is also the past!

Needless to say, time loops pose a serious problem for our neat Kantian
explanation of time-direction and causality.  If the future is the past,
which way does our experience go?  How can there be a diary today about
the future?  In brief, {\it what happens}?  If what happens is what could be
experienced by someone involved in the happening, what would such
experiences in a time loop be like?  And what would it mean for
experience itself to happen?  Before getting further into this quagmire,
let's briefly check in with the authorities to see whether time loops
are anything more than wild science fiction.

We'll start with Stephen Hawking.  Here is a recent item from the New
York Sunday Times:

``October 1, 1995.  In a U-turn that has sent shock waves through the
universe, Professor Stephen Hawking, Britain's leading cosmic physicist,
has accepted the possibility of time travel.  Having ridiculed the
concept for years, Hawking now says that it is not just a possibility
but one on which the government should spend money."

Hawking gives a quick summary of the reasons for his U-turn in the
preface to a new book called {\it The Physics of Star Trek} by astronomer
Lawrence Krauss\cite{Krauss96}.  First he addresses the problem of faster-than-light
travel, a staple of space opera despite its being forbidden by special
relativity, and concludes that ``Fortunately, Einstein's general theory
of relativity allows the possibility for a way around this difficulty."
He then turns to time travel:  ``One of the consequences of rapid
interstellar travel is that one could also travel backwards in time.
Imagine the outcry about the waste of taxpayer's money if it were known
that the National Science Foundation were supporting research on time
travel.  For that reason, scientists working in this field have to
disguise their real interest by using technical terms like `closed
timelike curves' that are code for time travel.  Nevertheless, today's
science fiction is often tomorrow's science fact."

One of the scientists working in this field is the super-string theorist
Michio Kaku.  Here are two brief excerpts from his recent book
{\it  Hyperspace}\cite{Kaku95}:

``In June 1988, three physicists (Kip Thorne and Michale Morris at the
California Institute of Technology and Ulvi Yurtsever at the University
of Michigan) made the first serious proposal for a time machine.  They
convinced the editors of Physical Review Letters, one of the most
distinguished publications in the world, that their work merited serious
consideration." ....

``If time travel is possible, then the laws of causality crumble.  In
fact, all of history as we know it might collapse as well."

Whatever may be the prospects for a time machine, and we personally don't
think they are very good, the conceptual problem posed by time loops in
outer space is problem enough in itself.  Even if we remain comfortably
at home in our causal near-Euclidean space-time and never venture near a
cosmic worm-hole, we still live in the same universe with such
aberrations.  In the larger picture, their time is of a piece with our
own, and thus the question still remains:  what happens there?  Since
this question can't be answered by a narrative, what kind of answer
should we look for?

Here's another brief quote from Kaku.  ``However, the concentrations of
matter-energy necessary to bend time backward are so vast that general
relativity breaks down and quantum corrections begin to dominate over
relativity."\cite{Kaku95} A complementary point is made in an article called
``The Quantum Physics of Time Travel" by David Deutch and Michael
Lockwood in the Scientific American, March 1994\cite{Deutch94}:  
``Quantum mechanics may
necessitate the presence of closed time-like curves.  These, while hard
to find on large scales, may well be plentiful at the submicroscopic
scale, where the effects of quantum mechanics predominate."

These are very interesting points, since, even without time loops,
quantum mechanics on the microscopic scale has always given a lot of
trouble to those who ask ``What happens there?".  It seems that we are
bracketed by temporal and causal weirdness from both the very small and
the very large.  The realm of causality and process, or at any rate due
process, appears to be confined to a middle scale, like sentient life
itself.  But are we so perfectly balanced in this middle that the
temporal weirdness at the extremes has no bearing at all on our
situation?

Not so long ago we seemed to be bracketed by weirdness of another kind:
Though pebbles are round and stars and planets are round, the Earth we
live on is flat.  Today we forget how incredibly hard it was for our
ancestors to give up their flat Earth.  The ancients managed to live
precariously on a round Earth for almost nine centuries, but by the
fourth century AD the Earth had become flat again, and flat even for
educated people who knew their Aristotle, Ptolemy and Aristarchus.
Curiously enough, it wasn't flattened by the barbarian invaders, but by
learned scholars who were well grounded in Alexandrian mechanical
technology which they used (or misused) to construct ingenious models of
the cosmos as a planetarium \cite{Dryer57}.

What drove our educated ancestors to such absurdities?  Biblical faith
is the usual suspect, but we think animal faith is the real culprit.
Animal faith says that up is up and down is down, and never the twain
shall meet.  UP and DOWN are Kantian-Darwinian categories, so deeply
entrenched in our psyches, our language and our genes that they pervade
all thought, as shown by expressions like getting high, feeling let
down, being well grounded, etc. etc.  The trouble with a round Earth is
not that we can't imagine its shape but that we can't imagine a universe
where unsupported things don't fall, and falling means going down.

The Aristotelian and Medieval solution to the problem of falling was to
have everything fall towards the center of the universe, which was
located in the center of the Earth.  But after the Copernican
revolution, the concept of falling once again became quite mysterious.
Which way do things fall when they are well away from Earth, and why?
It was Newton who solved this mystery by realizing that falling is not
something that happens to objects in isolation but is a relationship
between two or more objects.  With this insight, falling was subsumed
into another and very different concept, that of mutual attraction.

Past and future are even more ingrained in our genes, psyches, and
language than up and down, so quantum superposition and black holes are
even greater heresies for animal faith than a round Earth wandering
around the Sun.  Today it's not falling that is mysterious but
{\it happening}.  Does Newton's solution offer us any help in making it less
so?  Is it conceivable that we could subsume the concept of happening
into some more general concept?  If one is prepared to accept Kaku's
reading of the current astronomical
evidence, this larger concept can't have an absolute temporal arrow
anymore than Newton's concept could have an absolute arrow of up and
down.  The past and future have become poles in a relationship.  But a
relationship between what and what?  And described how?  We'll return to
this in a moment.

We agreed with Kant that {\it happening} is an empirical concept.  We then
simplified the meaning of {\it empirical} into being able to keep a diary of
observed events.  Let's now concentrate on the concept of a {\it diary}.

To keep a diary, you start out with a stack of white paper and write
down your experiences in the order in which they occur.  As we noted
above, each time you write something down, you narrow the set of
possibilities for what could end up on the paper, i.e. you take away a
subset of {\it cases}.  Each item in your diary corresponds to a certain set
of cases taken away, so the act of writing produces a sequence -C1, -C2,
-C3, ... of such departing case sets.

This account of writing is incomplete, however, since it only shows
cases departing, not cases arriving.  Our description of your diary
began after you had already acquired a stack of paper.  And yet,
properly speaking, your first act of diary keeping was to acquire that
stack of paper, since it is blank paper that {\it supplies} the set of cases
from which the members of C1, C2 , C3 etc. are taken away when you
write.  Furthermore, this first act may well be repeated, since you may
run out of paper before you are done.  Thus a more complete description
of your diary writing would go like this:  +C1, -C2, -C3, +C4, -C5 etc.,
where a +C is a set of cases added to those at hand when you get a new
sheet of paper, and a -C is a set of cases taken away when paper meets
pencil.

Now let's address the question:  what would it mean for a record of
events to be reversed in time?  What comes to mind is a movie run
backward:  spilled salt flying from the floor into its shaker, the
Phoenix rising from its ashes etc.  However, simply running the frames
of a movie backwards does not reverse the crucial order of events that
creates each frame, namely, light going from object to camera and then
leaving its imprint on the blank film.  A backwards movie is only a
forward record read backwards; it's not the record of backwards events.

Let's approach our question on the more fundamental level of information
and entropy, which is to say, on the level of departing and arriving
cases.  Once we take this course, its answer is immediate:  Time
reversal turns departing cases into arriving cases and vice versa.  What
goes backwards is not the order of movie frames or some other kind of
``state descriptions" but rather the order of the case sets C1, C2, C3
etc., and this is accompanied by a reversal of their signs.  Thus our
diary above would go +C5, -C4, +C3, +C2, -C1.   

Reversing past and future turns writing into paper and paper into
writing!

Time reversal is indeed very simple, but it's also pretty weird.  Here's
another bit of weirdness.  Seen forward in time, you take a piece of
paper supplying 100 cases and write something on it that takes away 50
of them.  Seen backward, you take a piece of paper supplying 50 cases
and take away 100 of them!  What is the meaning of the ``remaining" minus
50 cases?  It's certainly nothing familiar.  However, this example shows
that negative cases, and thus negative probabilities, can't be avoided
in describing time-reversal, and if the speculative cosmologists and
astrophysicists are right about
the circular warping of time, we will somehow have to make our peace
with them.

In short, the generalized concept of happening we have been looking for,
whatever else it may require, does require a generalized theory of
probability in which we can countenance debts as well as surpluses of
cases.      

In Chapter 1 we mentioned that negative cases play havoc with
logic.  Let's very briefly see how this happens, since the mathematical
``illogic" of negative cases is actually a useful tool. A more extended
discussion of the following points is given in 
``Boolean Geometry'', where the ``Boolean cube'' is pictured 
\cite{Etter96}.    

Define the
logic of a set of cases to be the set of all of its subsets.  Consider
the logic of a set of three cases.  It has eight members, which can be
arranged in a natural way as the vertices of a cube, called the {\it  Boolean
cube}.  If we stand the Boolean cube on the vertex representing the null
set 0, the top vertex becomes the universal set 1 containing all three
members, i.e. the principle axis containing 0 and 1 is vertical.  This
same construction can be carried out with n cases to give an
n-dimensional hypercube, which is also called a Boolean cube.

Ascending an edge of a Boolean cube (or hypercube) represents adding a
case to the lower vertex, where parallel edges always add the same case.
It follows that going down an edge represents taking away a case.  Since
writing a diary means successively adding and taking away cases, we can
represent the course of writing by an edge path that starts with the
vertex representing the initial case set.

Suppose now that we stand the Boolean cube upside down, putting 1 at the
bottom and 0 at the top.  Each step in the diary edge path then reverses
its sign, so this transformation of the cube represents time-reversal.
It also has a logical meaning, which is negation; more exactly, a
self-reflection of the cube in the vertical direction maps each vertex
into its negation.

Turning the cube upside down is only one of its many geometric
symmetries\cite{Etter96}.  We can classify these according to how far
they rotate the 0-1 arrow from its normal vertical orientation, the
extreme being 180 degrees, as in negation.  The intermediate angles
represent partial time reversal.  We can imagine the movable 0-1 arrow
as a throttle that ranges from full speed ahead in time when it is
pointing up to full speed astern when it is pointing down.  If the
throttle is exactly halfway between, the diary goes ``sideways" in time,
so-to-speak (this can only happen in a hypercube with an even number of
dimensions.)  Among such sideways transformations are square roots of
negation, which we'll come back to in a minute.

It was our aim to come up with a concept of diary that is invariant
under time reversal.  The most natural interpretation of its invariant
structure is a path on the Boolean cube, now understood as a purely
geometric structure without an intrinsic up-down direction.  This
conception relativizes the logic of the diary, since the operations AND,
OR etc. depend on the setting of the time throttle.  Recall in Chapter 1
that we noted the weird logic of interference, where A is possible and B
is possible but (A OR B) is impossible.  Relativizing logic explains
this puzzling situation:  The particle that ``interferes with itself" is
in an eigenstate of an observable for which OR has a different meaning
than it does in the frame of the classical apparatus.

Back to the square root of negation.  When we base probability theory on
Boolean geometry, this transformation gives us imaginary probabilities,
just as negation gives us negative probabilities; link theory turns
these into the imaginary and negative amplitudes of quantum mechanics.

Let us now go out on a limb and conjecture that the direction ``sideways
in time" is a direction in space.  This could mean that the square root
of negation is also behind the i of Minkowski space (see 
Section 3.8 for more on this topic.)  If so, it could be crucial for
unifying our theories of space and matter.

Let us go even further out on that limb.  First, We'll call on Pauli for
a bit of support.  Towards the end of his life Pauli had a mystical
vision of the quantum i as the key to unifying physics with psychology
\cite{Erkelens91}.  It turns out that complex quantum mechanics is
mathematically equivalent to a real quantum mechanics (see 
Section 3.6) in which every object contains a particular two-state
object, which has been called the Janus particle \cite{Etter95}, whose
state is unmeasurable.  TE's conjecture is that choosing between the two
states of the Janus particle corresponds to choosing between the
Heisenberg and Schrodinger representations.  The fact that the absolute
phase of amplitude is unobservable then means that quantum matter in
itself (i.e. unmeasured) is not only symmetrical with respect to time
direction but also with respect to {\it subject} and {\it object}.  The
differentiation of the world into particular pieces of matter,
particular minds, and particular space warps can thus be understood as
three different manifestations of the breaking of the primal symmetry
represented by i.

As we said before, logic is really out of bounds for this paper.  The
last few paragraphs must have seemed pretty cryptic, but we wanted to
give a bit of the flavor of things to come.  We'll return to such
speculations in the last section of Chapter 3, but until then our
business will be with matters closer to home.

Time reversal does more than just change the sign of the events in our
diary writing---it also reverses their order.  Now it's reasonable to
formalize the process of registering and recording information as an i-o
system.  As we saw, the laws of the underlying Markov process are
symmetrical in time, which means that we could also represent it by an
i-o system going backwards in time.  But what sort of system would
represent ``sideways" time?  Here we must turn from i-o systems to link
systems.  There turns out to be a whole spectrum of these ranging from
full speed ahead to full speed astern, and it is those exactly half-way
between that give quantum states (see Section 3.6).  Note that
the ``state throttle" here is very different from the ``phase throttle"
that rotates the Boolean cube, and indeed it need not even involve
negative probabilities.

We asked above whether {\it happening} might resemble {\it falling} in being a
special case of a more general situation analogous to motion in a
gravitational field.  As we'll see, the mathematics of the state
throttle supports this analogy.  We should speak of rising, however,
rather than falling, in the sense that events {\it arise} from the ``ground" of
past conditions.  With the state throttle in full reverse, events arise
backward in time from the ground of future conditions.  In the more
general case, events are ``repelled" by both past and future, like helium
balloons between two massive bodies.  At half throttle, when the
balloons are exactly balanced in between, we get quantum mechanics.

We've come a long way from our starting place, which was to respond to
Hume's challenge to find causality in the ``bare order of correlation".
In effect, Hume's problem effectively disappeared, giving way to others
more puzzling and serious.  The problem with Hume's problem is that it
assumes we are simply given a world where things happen.  This is indeed
true in everyday life, but when we start to think about what happens in
time loops or atomic nuclei, we find that it's the concept of happening
itself that is the problem.  Causality, time direction, and history all
seem to be bound up together in a way that makes it hard to say which is
the more fundamental.  This tightly knit complex of ideas is a familiar
one, and we have made some progress in analyzing it.  But we have also
discovered that it doesn't transfer very well to either the domain of
the very large or of the very small.  To make sense of modern physics
seems to require that we give up not only causality and history but even
logic.  The next chapter will lay out in a more systematic way the kind
of mathematical theory that must take their place.

\section{PROCESS AND SYSTEM: THE \hfill\break MATHEMATICS}

The main goal of this paper is to show that basic quantum mechanics can
best be understood as a branch of the theory of probability.  For this
to happen, probability theory must be expanded in two ways:  First and
foremost, probabilities must be allowed to go negative.  Second, and
this is more a matter of technique than a change in our assumptions, the
composition of transition matrices must be subsumed into a more general
method of composition based on the linking of variables.  We'll see that
{\it in this expanded theory}, probability amplitude is not some new kind of
``fluid" or physical field, but simply probability itself.  By regarding
it as such, we turn some of the most striking basic discoveries of
modern physics into pure mathematics, thereby freeing them from their
purely physical context to become tools available to every branch of
science.  Just how much more of physics as we know it will follow in
this course remains to be seen.

\subsection{Some Basic Concepts of Standard Probability Theory:}

{\it Event, sample space, case, variable, range, joint variable, logic, atom,
probability measure, probability distribution, random variable,
summation theorem, marginal, stochastic process, independence,
conditional probability, conditional independence, separability, Markov
property, Markov process.}

\noindent
{\bf Event}.  Following Feller\cite{Feller50} we'll take {\it event} to be our basic undefined
concept.  We'll use the word in pretty much the ordinary sense.  Like
most English nouns, it can be either definite or
indefinite.

\noindent
{\bf Sample space, case}:  A {\it sample space} is a set of mutually exclusive
possibilities called {\it cases}.  We say that an event E {\it belongs} to a sample
space L if for every case of L, it makes sense to ask whether E is true
or false.  Another word for event is {\it proposition}.  Examples:  The event
{\it Heads} belongs to the sample space whose cases are the possibilities
Heads and Tails.  The event {\it Seven} belongs to the sample space whose
cases are the 36 possible throws of a pair of dice.

\noindent
{\bf Variable}:  The term will be used in the scientist's sense, meaning
something that can vary, rather than in the logician's sense of a
place-holder for a term.  A variable x is said to belong to a sample
space L if for any value k of x, the event x=k belongs to L. For
example, let L be the set of all cases for ten coin tosses, and let x be
the number of heads; then x=1, x=2, x=3 ... x=10, are events in L and x
is a variable in L. More abstractly, a variable is simply a function on
L; when we regard x in this way, the proper notation is x(l).  The set
of all possible values of a variable is called its range.

\noindent
{\bf Joint variable x ${\cal J}$ y}:  Though variables are often numerical, we'll place
no restrictions on what kind of things their values can be.  For
instance, a variable can range over all of the joint values of several
other variables.  The notation x ${\cal J}$ y will refer to a variable that ranges
over all joint values of x and y.

There is an ambiguity in the term ``joint value", which is whether or not
it involves the order of things joined.  Is x ${\cal J}$ y the same variable as
y ${\cal J}$ x?  And is (x ${\cal J}$ y) ${\cal J}$ z the same variable as 
x ${\cal J}$ (y ${\cal J}$ z)?  Actually, there are
two concepts of joining, which in everyday language we would render as
``x and y" and ``x, then y".  The first and simplest, which is
unstructured joining, is commutative and associative, so we can write
x ${\cal J}$ y ${\cal J}$ z ${\cal J}$ ...etc... without parentheses 
and with the variables in any order.
The second, which is the one we'll use here, takes joint values to be
{\it ordered} n-tuples; it has the decisive advantage over the first that
variables of the same type are in natural 1-1 correspondence.  There's
actually a third concept called {\it list structured} joining that is very
useful in constructing complicated link diagrams, but it belongs to a
more advanced and specialized chapter of our theory.

\noindent
{\bf Logic, 1 and 0, generating}:  Events can be combined with AND, OR and
NOT, and a set of events closed under these operations will be called a
{\it logic} (this use of the word `logic' comes from von Neumann).  We'll
abbreviate AND by ``\& '' and NOT by ``$\sim$''. A logic is of course a Boolean
algebra.  The {\it null} element of a logic, abbreviated 0, is the element
A\& $\sim$A, which is the same for any A. The {\it universal} element, abbreviated 1,
is the element $\sim$0 = (A OR $\sim$A).  
Given any set K of events, the set of
all events that result from applying AND, OR and NOT to the members of
K is a logic, and it will be called the logic {\it generated} by K. Given a
variable x, the logic generated by all events of the form x=k, where k
is a value of x, is called the logic of x, and the members of that logic
are called events in x. More generally, the logic of x, y, z ... is the
logic generated by all events of the form x=i, y=j, z=k etc., and the
members of that logic are called events {\it in} x, y, z etc..

\noindent
{\bf Atom}.  The atoms of a logic are the cases of that logic.  To put it
another way, an {\it atom} of a logic is defined as an event that cannot be
further refined by any other event of the logic, i.e.  A is an atom if,
for any E in the logic, either A\&E = A or A\&E = 0. Atoms are mutually
exclusive possibilities, and the set of all atoms is exhaustive within
the logic.  In the logic of the variable x, the atoms are the events of
the form x=k for all k, while in the logic of x, y, z... the atoms are
the events of the form (x=i)\&(y=j)\&(z=k)... for all i, j, k....  If the
number of atoms is finite, then the subsets of atoms uniquely correspond
to the events in the logic.  The cases of the sample space are the atoms
of its logic, so atom and case are basically synonymous terms.  However,
we'll generally reserve the word `case' for atoms that we count, such as
the atoms of the sample space.

\noindent
{\bf Probability}.  Underlying all our more sophisticated ideas about
probability is Pascal's classic definition of the probability p(E) of an
event E as the number of cases favorable to E divided by the total
number of cases.  If the cases of a sample space can be regarded as
equally probable, then this definition defines a so-called {\it probability
measure} on the events of that sample space, satisfying the following
three axioms:

\noindent
1) It's additive:  If E1 contradicts E2 then p(E1 OR E2) = p(E1)+p(E2)

\noindent
2) It's non-negative; for all E, p(E) $\geq$ 0.

\noindent
3) It's normal, i.e. p(1) = 1.

For now we are dealing with {\it classical} probabilities, which are never
negative.  For these we can always imagine an underlying sample space
with equiprobable cases that define p(E) according to Pascal's
definition.  In dealing with negative probabilities, the notion of an
underlying case set must be generalized so that it can contain both
positive and negative cases. (The notion of negative set membership has
been carefully investigated by mathematicians, and Blizard has shown
that the Zermello-Fraenkel axioms can be generalized to allow for it
\cite{Blizard90}.)  In this more general form, the case set has a very
important role to play in link theory, since it enables us to
distinguish among several quite different conceptions of independence;
the details here will be found in future papers.

\noindent
{\bf Probability distribution on x}:  A function defined on the range of x
which for each k in that range has the value p(x=k).  The standard
notation for this function is p(x), which can be very confusing, since
it conflates the scientist's and logician's two very different notions
of variable.  However, it's such a well-established tradition that we'll
have to live with it; when you get confused, just remember to replace x
by something that looks like a sentence.

\noindent
{\bf Stochastic, or random variable}:  A variable on which is given a
probability distribution.  The qualifiers `stochastic' and `random' are
only needed when a variable must be distinguished from variables of
other kinds, like free variables.  Since our results have nothing to do
with randomness, 'stochastic' is the preferred qualifier.

\noindent
{\bf Joint probability distribution} on variables x, y, z..., written
p(x,y,z..) is defined as the probability distribution on the variable
x\&y\&z...

\noindent
{\bf Summation theorem and marginals}:  The summation theorem says that given
a joint probability distribution on variables x and y, we can find the
probability distribution on x alone by summing over all values of y;
i.e.. p(x=i) = $\Sigma_{j\in y}$p((x=i)\&(y=j)), or, in our abbreviated notation,
p(x) = $\Sigma_y$p(x,y).  
This is a theorem we'll use repeatedly.  A corollary is
that a joint probability distribution uniquely defines the joint
probability distribution on any subset of its variables; such partial
joint probability distributions are called {\it marginals}.

\noindent
{\bf Stochastic process}.  Define a {\it stochastic process}, 
or {\it process} for short,
as a set of variables on which is given a joint probability
distribution.  In the usual definition it is also assumed that these
{\it stochastic variables} involve events which can be ordered in time.  This
won't do for our present purposes, though, since, we need a new concept
of process general enough to apply to time loops.  We can start out by
thinking of process variables as having at least approximate locations
in the space-time of current physics.  However, since we aim to
eventually construct space-time out of probabilities, this is only a
temporary expedient.

{\bf Independence}.  Events E, F, G,...  are called independent if 
p(E\&F\&G...) = p(E)p(F)p(G)....  Note that for a set of events to be independent,
it's not enough that every pair of events in that set be independent;
independence is a joint condition of all the events.  Two variables are
called independent if events in the first are independent of events in
the second.  A set of variables are called independent if events in
different variables are independent.   This definition pretty much
captures our intuitive concept of independence as lack of correlation,
although there is one thing about it that is not so intuitive:  Suppose
that either p(E)=1 or p(E)=0.  Then for any F, p(E\&F) is either p(F) or
0; in either case it is p(E)p(F).  Such a ``definite" E is independent of
all events, including itself!  This oddity will turn out to be important
for the concept of probability space (see Section 3.8).

Given a finite sample space of equiprobable cases, there is another way
to define the independence of E and F, which is that the sample space
can be arranged as a rectangular set in which the cases of E are a
vertical stripe and the cases of F are a horizontal stripe.  For
ordinary probability theory in which cases always count positively, the
two definitions are equivalent.  However, with both negative and
positive cases, the two diverge, since p(E) can be 0 even though its
cases set is not ``rectangular" with respect to that of F. It turns out
that there is even a third definition of independence which applies to
imaginary and complex probabilities (see Section 3.6) when these are
defined in terms of real probabilities.  The above product rule is the
definition we'll adopt here, however.

\noindent
{\bf Conditional probability}.  The conditional probability of E given F,
written p(E $\vert$ F), is defined as p(E\&F)/p(F).

Since a conditional probability results from dividing by p(F), it is
only well-defined if p(F) is not 0. Negative probabilities are
constantly producing 0's in weird places, so it's best to avoid
conditional probabilities if there's an alternative; fortunately, there
usually is.  

\noindent
{\bf Condition}: An event C that is assumed to be true, thereby
reducing the sample spaces to the set of those cases favorable to it.
Axioms for a process are often given as conditions on a ``free" process
whose variables are independent.  A very important kind of condition for
our present work is the {\it link}, which is an event of the form x=y.  The
probability of E in the reduced sample space conditioned by C is 
p(E $\vert$ C) in the unconditioned sample space.  

\noindent
{\bf Conditional independence}:  Events E
and F are called {\it conditionally independent} given C if they are
independent in the sample space conditioned by C. The usual algebraic
formulation of this relationship is p(E\&F $\vert$ C) = 
p(E $\vert$ C)p(F $\vert$ C).  More
generally, a set of events E1, E2, E3... is conditionally independent
given C if p(E1\&E2\&E3... $\vert$ C) = p(E1$\vert$C)p(E2$\vert$C)p(E3$\vert$C)...    

As we noted in
Chapter 2, these equations becomes meaningless when the probability of
the condition is 0. The following definition is equivalent to
conditional independence when p(B) is not 0, but also makes sense when
it is 0:  

\noindent
{\bf Separability}.  Events A and C are called separable by event B
if p(A\&B\&C)p(B) = p(A\&B)p(B\&C).  This definition comes from
noting that it can be obtained by multiplying the conditional 
independence condition, p(A \& C $\vert$ B)= p(A $\vert$ B)p(C $\vert$ B), on 
both sides by p(B)$^2$.  The equation defining
the separability of n events by B is derived by multiplying both sides
of the appropriate equation for conditional independence by p(B)$^n$.

We say that variable y separates variables x and z if p(x,y,z)p(y) =
p(x,y)p(y,z). which is an elliptical way of saying that if for any
values i, j, and k of variables x, y and z we have
p((x=i)\&(y=j)\&(z=k))p(y=j) = p((x=i)\&(y=j))p((y=j)\&(z=k)).

Much of our work will be with Markov processes.  Feller\cite{Feller50} begins his
chapter on Markov processes with the concept of Markov chain, which he
defines as a series of trials in which the probabilities at each trial
depend on the outcome of the previous trial, the rule of dependence
being the same for all trials.  He then goes on to say that a Markov
chain is characterized by a matrix of conditional probabilities, called
the {\it transition matrix}, together with an initial ``state vector" giving
the probability distribution at the initial trial.

There are problems with this definition.  Part of what motivates it is
the need to separate out the {\it law} of the process, given by the transition
matrix, from the {\it boundary condition} on the process given by the initial
state vector.  And yet there is no reason why the initial vector should
not contain 0's, for which conditional probabilities are meaningless.
Thus the Markovian law of the process, which makes perfect intuitive
sense whatever the initial state, cannot properly be described by a
matrix of conditional probabilities.  The so-called transition matrix
must be defined in some other way.

The remedy for this and related confusions is one we have discussed at
length in Chapters 1 and 2; this remedy is to clearly distinguish between the
process itself and its representation as a {\it system}.  Feller began by
describing a process and then switched over to describing it as a
system, without noticing the change.  Section 3.6 will be devoted to
Markov systems, so here we'll define only the process.  First a more
general concept, which we've already met in Chapter 2:

\noindent
{\bf Markov property}:  Three events A, B and C are said to have the Markov
property if p(C $\vert$ A\&B) = p(C $\vert$ B).  Multiplying both sides by p(A\&B)p(B), we
see that this is simply separation, i.e. it says that B separates A from
C

\noindent
{\bf Markov process}:  A stochastic process whose variables can be arranged in
a sequence such that each variable separates the variables before it
from those after it.

\noindent
{\bf Markov transition matrix}:  Abstractly, a matrix in which the sum of every
column is 1 (just to confuse matters, such matrices are sometimes called
{\it stochastic matrices}.)  More concretely, the matrix of conditional
probabilities from variable x(t) to variable x(t+1) in a Markov process.

\subsection{Processes}

\noindent
{\it Primary variables, secondary variables, join, global variable,
independent parts.}

Here are some useful technical terms:

\noindent
{\bf Primary variables}:  In Section 3.1, a process was defined as a set of
random variables on which we are given a joint probability distribution;
these variables will be called the {\it primary} variables of the process.  We
will assume that the primary variables are given in a certain linear
order, which may or may not be related to the probability distribution.

\noindent
{\bf Secondary variables}:  A secondary variable is a joint variable in a set
of primary variables, called its members.  We will assume that the
members always occur in their natural order, so there is one and only
one secondary variable for every subset of primary variables.  We'll
speak of secondary variables as also belonging to the process.  Both
primary and secondary variables will be denoted by small letters.

\noindent
{\bf Variables}:  Unless otherwise stated, the word `variable' will refer to
either a primary or secondary variable of a process.

\noindent
{\bf White variable}:  A variable on which the probability distribution is
uniform.

\noindent
{\bf Global variable}:  The secondary variable whose membership includes all
primary variables.  The process itself will be denoted by capitalizing
the global variable, e.g., W is the process whose global variable is w.

\noindent
{\bf Independent part, component}:  A sub-process U of W is called an
independent part of W, or a {\it component} of W, if the primary variables of
U are independent of all other primary variables of W. The primary
variables of U are also called its {\it members}.

\noindent
{\bf Prime}:  A process having no independent parts is called {\it
prime}.

\noindent
{\bf Prime factorization theorem}.  The set of prime components of a process
is unique.

One often hears that the whole is more than the sum of its parts.  What
is generally meant is that the whole is more than the mere heap of its
parts, which is true unless the whole happens to be a heap of parts.  A
process is indeed the heap of its prime parts, which are in fact
components, in the sense that the description of each does not require
that we take into account the individual peculiarities of the others.
However, we saw that one of the surprises of negative and imaginary
probability theory is that there are several quite different definitions
of independence, and thus the members of a heap share, as it were, the
``nationality" of their heap, which involves the particular kind of
independence that its members have from each other.

\subsection{Links and Cuts}

If you have been dozing through these definitions, that's OK---you can
always go back to them.  But now is the time to wake up.  Pay attention!
Here comes the key idea of the whole paper:

\noindent
{\bf Link}:  Given a process W containing variables x and y, we {\it
link} x to y in
two steps.

\noindent
Step 1:  Apply the condition x=y to W to obtain a new {\it conditioned}
process W' with the same variables as W but with a new joint probability
distribution W' in which the probability p'(E) of an event E is p(E$\vert$x=y)
in W.

\noindent
Step 2:  Then modify W' by dropping the duplicate variable y.

We can think of a link as an operator L(W,x,y) on W, x and y that
produces a new process W', in which the probability of an event E is its
probability in W conditioned by the event x=y, and in which the variable
y is dropped.  We'll write W$\vert$(x=y) for this operation, i.e.  W' =
W$\vert$(x=y); more generally, W$\vert$LMN will mean applying links L, M and N to
process W.

There is nothing in the definition of link that says x and y must be
different variables; W$\vert$(x=x) is a perfectly legitimate operator.  What,
then, does it do?  It does nothing to the probabilities, since the
condition x=x is already in effect.  It does, however, drop the second
variable, namely x. To link a variable to itself simply means to drop it
from W.

x and y can be either primary or secondary variables.  If they are
secondary, however, they must contain corresponding members that are
also implicitly linked.  Thus if x = x1\&x2 then y must be of the form
y1\&y2, and the condition x = y implies also that x1 = y1 and x2 = y2.
This means that when y is dropped, we must also drop all of its primary
members, and with them all other secondary variables containing any of
these members, which are no longer needed, since they are duplicated by
secondary variables with corresponding members in x. Dropping a variable
is a bit more complicated than the name suggests.

The converse of linking could be called 'de-linking', but the term
doesn't really fit, since it implies that there are two things to be
de-linked, whereas in fact there is only one.  To restore the original
W, we must first turn x in two equal variables, and then remove the
condition that equates the two.  We'll call this operation:

\noindent
{\bf Cutting}:  To {\it cut} the process W' at x means to produce a new process W
containing a new variable y such that W' = W$\vert$(x=y).

Unlike linking, cutting is not an operator, since there are an infinite
number of ways to cut W' at x.

\noindent
{\bf Equivalent cuts}:  Cuts on the same variable whose linking produces the
same process.

\subsection{States}

{\it Link state, density matrix, state of a variable, trace, propositions,
projections, Born's rule, pure states, mixed states.}

We now come to another key concept.

\noindent
{\bf State of a link}:  Given a link W$\vert$x=y, the state of that link is defined
as the joint probability distribution on x and y in W.

\noindent
{\bf Equivalent states}:  States that result from equivalent cuts.

\noindent
{\bf Density matrix}:  The joint probabilities on x and y arranged in a matrix
with x as the horizontal index and y as the vertical.  Following {\it
one} quantum tradition, we'll use the terms `state' and `density matrix'
interchangeably.

\noindent
{\bf State of a variable x}:  The density matrix of the link x=x.  Notice that
the off-diagonal elements of this matrix are all 0, since it's
impossible for its two indices to differ, and impossibility has
probability 0!  A more rigorous definition of the state of a variable is
the diagonal matrix whose indices range over the range of that variable,
and whose diagonal is the probability distribution on that variable.
For instance, if x ranges from 1 to 3, then its state is the 3 by 3
matrix M$_{ij}$ such that M$_{11}$ is the probability that x=1,
M$_{22}$ the
probability that x=2, M$_{33}$ the probability that x=3, and all other
entries are 0.

We'll find that the state of x encompasses all of the classical meanings
of the word `state', including Markov states and deterministic states
like those of a computer.  A Markov state is usually represented by a
vector; this traditional vector can now be defined as the diagonal of
the state matrix of the time-dependent random variable of the process
(We'll see the details of this later).

\noindent
{\bf Trace}.  The trace of a matrix S, written tr(S), is defined as the sum of
its diagonal entries.

\noindent
{\bf Important trace theorem}:  tr(AB) = tr(BA).  Important corollary:  tr(A)
= tr(TAT$^{-1}$), i.e. the trace of a matrix is invariant under all linear
transformations.

\noindent
{\bf Proposition}:  Literally, a synonym for event.  (We are borrowing the term
from von Neumann, who applied it to quantum variables.)  There is a
difference in connotation, though.  Events are often described by noun
phrases, whereas one always represents propositions by sentences.  We
can think of propositions as sentences in the predicate calculus whose
free variables are assigned to process variables.

\noindent
{\bf Predicate}:  A ``propositional function" whose free variables are
unassigned.  A predicate becomes a proposition when all of its
variables are either assigned or quantified.  A link, for instance, is a
proposition resulting from the assignment of the equality predicate to a
pair of variables.

\noindent
{\bf Projection}:  In linear algebra, a projection P is defined as a linear
operator satisfying the idempotent law PP = P. It can be shown that we
can always find a basis in which such a P is represented by a diagonal
bit matrix.  For now we'll confine ourselves to diagonal projections,
i.e. a {\it projection} will be defined as a matrix in which all off-diagonal
elements are 0 and the diagonal elements are either 0 or 1.

Von Neumann showed that propositions about the measurement of a quantum
observable x can be represented by projections that commute with the
eigenvalues of that observable.  This led him to an elegant
generalization of the Born probability rule:  If P is a projection
representing a certain proposition, and D is a density matrix
representing a quantum state, then, given D, the Born probability of
that proposition is tr(PD).

\noindent
{\bf Representing predicates by projections}:  If P(x) is a proposition about
x, define the {\it characteristic function} of P(x) to be the bit-valued
function B(x) which is 1 or 0 corresponding to whether P(x) is true or
false for x. The diagonal matrix P having B(x) as its diagonal is the
projection that we'll use to represent the proposition P(x).  We'll
normally use the same letter P for both the predicate and the
projection, and when the predicate is assigned, for the resulting
proposition; it's usually clear from the context which of these things
we mean.

If we multiply the projection P by the state matrix D of x, we clearly
knock out all cases for which the predicate P is false, so the sum of
the remaining entries in D is the probability of proposition P, i.e.
p(P) = tr(PD).  This makes Born's probability rule start to look a bit
less mysterious.  However, what really demystifies Born's rule is that
it holds not just for the state of a variable but for any link state.
More exactly:

\noindent
{\bf Generalized Born's rule}.  Let S be the link state of x=y in process W,
and let P be any proposition about x. Then the probability of P in the
conditioned process W$\vert$x=y is tr(PS)/tr(S).

\noindent
Proof:  First note that the state S' of x in W$\vert$x=y is S with all
off-diagonal elements set to 0, normalized by dividing by p(x=y).  Since
P is diagonal, PS is the result of multiplying the rows of S by
corresponding diagonal elements of P. Thus the diagonal elements of S'
are simply those of S divided by p(x=y) = tr(S).  Since we just saw that
p(P) = tr(PS'), we conclude that p(P) = tr(PS).  QED.

Our result differs from von Neumann's by the factor tr(S).  Had we
defined a link state as S/tr(S), thus making tr(S) always 1, we would
have gotten von Neumann's result exactly.  This was our original
definition of state, and it works fines for classical and quantum
states.  However, we would like our definition of state to still make
sense even as we push our theory into new domains where, because of the
possible cancelation of plus and minus cases, the trace of S can be 0.
For this reason, We believe that our present unnormalized state is
preferable in the long run.

\noindent
{\bf Pure states}.  Given a link W$\vert$x=y, if x and y are independent in W, the
state of the link is called {\it pure}.

It follows directly from the definition of independence that the density
matrix of a pure state has the form $\vert$v$><$u$\vert$ in Dirac notation.  If
u is
white, the state will be called {\it causal}.  On the other hand, if u=v, the
state is quantum.  Causal and quantum overlap  if u and v are both
white., but in no other case.

\noindent
{\bf Pure causal state}:  The state of a link between independent variables,
of which the second is white.

\noindent
{\bf Causal trace theorem}:  If S is pure causal, then tr(S) = 1/n, where n is
the dimension of S.

\noindent
Proof: To say that w is white means that for any i, p(w=i) = 1/n.  Let
S be the state of link W$\vert$x=w.  Since S is pure, p((x=i)\&(w=i)) =
p(x=i)p(w=i) = (1/n)p(x=i).  But p((x=i)\&(w=i)) is S$_{ii}$, the i'th
diagonal term of S. Thus tr(S) = $\Sigma_i$ S$_{ii}$ = 
(1/n)$\Sigma_i$ p(x=i) = 1/n.

\noindent
{\bf White connection theorem}:  Roughly:  Pure causal links don't affect the
component on the ``output" side.  More exactly, if x is in a component A
that is independent of y, and y is white, then the link W$\vert$x=y doesn't
change the joint probability distribution on A.

\noindent
Proof:  Let z be the join of the variables of A other than x, i.e.  A =
X\&Z.  Then p(x,z,y) = p(x,z)p(y) = (1/n)p(x,z) in W. By the above trace
theorem, p(x=y) = 1/n.  Thus p(x,z,y$\vert$x=y), which, ignoring y, is the
joint probability distribution on x and z in W$\vert$x=y, is equal to
(1/n)p(x,z)/p(x=y) = p(x,z), which is the joint probability distribution
on x and z in W. QED.

This theorem is the key to mapping the engineer's input-output systems
into link theory.  Remember, we asked the question:  When an input is
wired to an output, how do you know which is which?  The answer is:  cut
the wire and see which loose end is unaffected by the cut.  This answer
can be amplified a bit.  If the wire is from box A to box B which is
otherwise unconnected to A, then cutting it not only leaves the output
of A unaffected, but all of A unaffected.  This is essentially the
content of the white connection theorem, if we regard the causal link as
a connecting wire.  The word ``causal" means what it says!

\noindent
{\bf Mixed states}:  Needless to say, pure states play a very important role
in our theory.  In quantum mechanics, impure states are called {\it
mixed}
states; this is because they can always be written as linear
combinations of pure states.  This turns out to be true quite generally
of link states, so we'll borrow the quantum term {\it mixed state} for any
state that is not pure.  Let us point out, though, that the
decomposition into pure states of a mixed classical state, such as one
might find in a computer program, is generally quite artificial, since
the pure components will seldom be classical and will often involve
negative probabilities.

\subsection{Link Systems}

{\it Disconnection, fundamental theorem of disconnection, link systems,
proper systems, predicates as components, relational composition,
abstract links and their states, contraction, product theorem, tensors,
public variables, systems, transition matrices, causal boxes and
systems.}

We now come to a simple but profound connection between purity and
separability which could be the key to unifying space and matter.  First
a definition:

\noindent
{\bf Disconnection}:  To {\it disconnect} x from z at y means to cut their process
at y in such a way that x\&y becomes independent of y'\&z , where y' is
the new variable produced by the cut.

\noindent
{\bf Fundamental theorem of disconnection}:  Let W be the join of three
variables x, y and z, i.e. w = x\&y\&z,, where p(y) is not 0 for any y.
Then x can be disconnected from z at y if and only if y separates x and
z. (Recall that separation means p(x,y,z)p(y) = p(x,y)p(y,z).)

\noindent
Proof:  

Suppose there exists a disconnection of W that produces a new
process W'.  Let y' be the new variable of W', i.e.  W = W'$\vert$(y=y').  To
avoid confusion, let's use p for probabilities in W and p' for
probabilities in W'.  Then in W' we have p'(x,y,y',z) = p'(x,y)p'(y',z).
This implies that, in W', x is independent of z for any values of y and
y'.  In particular, x is independent of z if y=k and y'=k.  Since this
is true for any k, it is consistent with the linking condition y=y',
which is what turns W' into W. This shows that y separates x and z in W.
Notice that going from disconnection to separation doesn't require the
assumption that p(y) is never 0.

Going backwards is a bit harder.  Suppose y separates x and z in W. We
need to construct a new W' containing x, y and z plus a new variable y'
such that x\&y is independent of y'\&z in W', and such that W = W'$\vert$(y=y').
Separability in W means that p(x,y,z)p(y) = p(x,y)p(y,z).  Let W' be a
process consisting of two independent parts W1' and W2', where W1'
contains x and y, and W2' contains y' and z. Let p'(x,y) = p(x,y).
We construct p'(y',z) as p(z$\vert$ y')/n where n is the number of
cases of z for which p(z$\vert$ y') is non-null. Then p(z$\vert$ y')
=p(z\&y)/p(y') where p(y') = $\Sigma_z$ p(z$\vert$ y'). Note that this
is where the requirement p(y') $\neq$ 0 is needed.w  

We must now show that under the
condition y'=y, W' is W, which is to say, p'(x,y,z) = p(x,y,z).  First,
we will use separability to separate out z in the joint probability
distribution of W.

\noindent
{\bf Lemma 1}: p(x,y,z) = p(x,y)p(z$\vert$y).  Proof:  Because p(y) is never 0, we
can freely speak of probabilities conditioned by y. Thus we can write
p(x,y,z) as p(x,z$\vert$y)p(y).  We can also write separability as
conditional independence, i.e. p(x,z$\vert$y) = p(x$\vert$y)p(z$\vert$y).  Combining the
two we get:  p(x,y,z) = p(x,y)p(y)p(z,y) = p(x,y)p(z$\vert$y).  QED

\noindent
{\bf Lemma 2}. The joint probability distribution on x, y, y' and z in W' is
p(x,y)p(z$\vert$y')/n.  Proof:  By the independence of W1' and W2', we have
p'(x,y,y',z) = p'(x,y)p'(y',z).  Since p'(x,y) was defined as p(x,y)
this becomes p(x,y)p'(y',z).  = p(x,y)p(z$\vert$y')/n, where n is the number
of rows in the transition matrix p(y'$\vert$z).  QED

Now let's link y and y'.  This involves two steps:  first we identify
the variables y and y', and then we normalize the resulting distribution
by dividing by p'(x=y).  We see by lemma 2 that the first step turns the
joint distribution on W' into p(x,y)p(z$\vert$y)/n.  Now p'(x=y) is the trace
of the state matrix of the link, which we can easily see is 1/n (the
steps showing this are spelled out in the proof of the causal link
chain theorem in the next section.)  Thus dividing by p(y=y') yields
p(x,y)p(z$\vert$y), which by lemma 1 is p(x,y,z).  Thus p'(x,y,z) = p(x,y,z).

QED.

The above disconnection theorem is what tells us how to mark a process
for dissection, so-to-speak.  First, we find a variable x, primary or
secondary, which separates the process into two conditionally
independent parts.  We then disconnect the process at x into two
unconditionally independent parts, and then repeat for these two parts
etc. until we can repeat no more, at which point we have arrived at a
prime factorization.  These prime factors, as such, constitute a mere
heap of independent parts.  The disconnection procedure, however, leaves
a record of past connections, a kind of wiring diagram, that can be used
to reconstitute the original process.

The question arises whether such a prime factorization is unique.  There
are actually two questions here:  first, whether the factors themselves
are unique as processes, to which the answer is definitely no (except in
the trivial case where the process has no factors except itself) and
second, whether the prime ``boundaries" are unique, or to put it another
way, whether every factoring procedure ends up with the same sets of
primary variables in its factors.  The second question has no simple yes
or no answer.  Often it is yes, but there are important cases (EPR, for
instance) where a seemingly arbitrary choice in making the first cut
will determine whether or not any further cuts are even possible.

\noindent
{\bf Link system}:  A pair of processes W and W' together with a sequence of
links on W that produce W'.

\noindent
{\bf Proper link system}:  A link system in which all links are between
different independent parts of W, and no two links involve the same
variable.

\noindent
{\bf Propriety theorem}:  In a proper system, the links commute, i.e. you can
get from W to W' by applying the links in any order.

\noindent
Proof:  Divide the linked variables into two sets X and Y, where X
contains the first variable of each link, and Y contains the second.
Let x be the join of variables in X, y the join of variables in Y. Then
the link x=y clearly encompasses the links of all the members of X and
Y, whose order is immaterial as long as they are properly matched.

The linking order can indeed make a difference if the linked pairs have
variables in common.  If you link y to z and then x to y, all three of
x, y and z will be tied together and called x. However, if you first
link x to y, y disappears, so you when then you try to link y to z,
there is no process variable to link.  This sort of thing cannot happen
in a proper system, of course.

But suppose you need to link all three of x, y and z together?  Indeed,
this is exactly what an engineer does all the time when he ``branches" an
output to go to several inputs.  Must the engineer then work with an
improper system?

The answer is no.  There are in fact several ways to branch properly.
One rather interesting way is to extend link theory by treating
predicates as ``abstract" components which, when linked to random
variables, become conditions on the process.  Thus to connect x, y and
z, we introduce the abstract component x'=y'=z' plus the three proper
links.  The resulting more general theory merges stochastic composition
with the predicate calculus, and may be of mathematical interest.
However, much of what it accomplishes can be done in other ways; for
instance, we can regard x'=y'=z' as a stochastic process with three
equal white variables, and the links x=x', y=y' and z=z' also connect x,
y and z. At any rate, abstract components won't concern us here.  With
one exception, that is; links themselves should definitely be regarded
as abstract components.  But they are the only ones we need for now.

Not having to worry about the order and grouping of links is such a plus
that we'll assume that all our systems are proper.  Under this
assumption, there is a very readable and natural diagrammatic notation
for link systems, which uses icons to represent component processes, and
arrows between them going from x's to y's to represent links.  Indeed,
if we draw and label the icons right, we can often avoid variable names
altogether, since the variables will be identified by their places.

Unfortunately, the practicalities of word processing, e-mail, etc. make
it expedient to have a linear backup notation.  Boxes are written thus:
A[y1,y2, .. ; x1, x2, ..], where the variables before the semicolon are
the ingoing arrows, those after, the outgoing.  A link is shown by an
equation, so A(y ; x) (x=x') B(x' ; z) links x in A to x' in B. If there
is no need to reconstruct the particular lost variables in W, we can
omit the equation and simply use the same letter in A and B, e.g.  A(y ;
x) B(x ; z).  Needless to say, diagrams are much easier to understand,
and it's often best to turn this linear notation into actual diagrams
rather than trying to decipher it as it is.  There are more powerful
diagrammatic notations than this for dealing with complex structures,
but they aren't needed for the simple examples we'll be studying here.
It's worth introducing one new concept from this extended notation,
though:

\noindent
{\bf Box}:  An icon that represents a component of a system, but that may also
be a container, implicit or explicit, for a more detailed system.  From
the outside a box is to be regarded simply as a process, but it's
permissible to look inside it for smaller components (think of the
transistors inside an integrated circuit 
or the electrons inside an atom.)  In the next
two sections we'll show all component icons as boxes, and sometimes refer
to them as such.

\noindent
{\bf Transformation}:  A two-variable sub-process T of a link system in which
both variables are linked to others.  If the variables of T are x' and
y, in that order, and if x=x' is the link to x and y=y' is the link to
y, then the link state (density matrix) of x=x' is called the state
{\it before} T, while the link state of y=y' is called the state {\it
after} T. T is
called {\it causal} if x' is white.

\noindent
{\bf Transformation product theorem}.  Roughly, to combine two successive
transformations, multiply their matrices.  More exactly, if T and U are
transformations with variables x',y and y',z, then linking y and y' and
then ignoring y produces the transformation nUT, where n is a
normalizer.

\noindent
Proof:  Combine the summation theorem with the definitions of linking
and matrix multiplication.

\noindent
{\bf Corollary}:  Equivalent cut theorem.  Suppose that cut C1 in W' produces
the process W1 with variables x and y. Define W2 as the process that
results from multiplying x by diagonal matrix D and y by D$^{-1}$.  Then the
cut W',W2 is equivalent to the cut W',W1.

\noindent
{\bf Corollary}:  A pure state with no 0's in its diagonal is equivalent to a
causal state.

\noindent
{\bf Corollary}.  Born's rule gives the same probabilities for equivalent
states.

\subsection{Markov Chains and Schrodinger's Equation}

{\it The rule of law, boundary conditions, Markov chains, generators,
time-reversed Markov chains, link chains, the link dynamical rule,
differential Markov chains, inverse Markov chains, double boundary
conditions, chains, sharp states, Schrodinger's equation, complex
amplitudes, the phase particle, the Janus particle.}

The ancient and Medieval worlds observed nature as carefully as we do,
hoping as we do to discover what remains fixed in the midst of change.
But science, as we know it, began when our focus shifted from changeless
{\it things} to the changeless laws of {\it change itself}.  The goal of ``natural
philosophy" then became to find those universal ``laws of causality" that
determine the future from the present.  The amazing success of Newtonian
mechanics convinced many people that this goal had actually been
attained, and so went the scientific consensus for 200 years.  But then
came quantum mechanics.

What most shocked people about quantum mechanics was its seeming break
with Newtonian determinism.  Chance, seemingly exiled to the limbo of
ancient superstition, had, like Napoleon returning from Elba, once again
forced itself upon the civilized world.  Could Chaos be far behind?  The
ideal goal of scientific theory had to retrench from predicting
certainties to predicting probabilities.  Our ultimate ``law of
causality" was an illusion; the best we could hope for was to find the
ultimate rule giving the best {\it probability distribution} on the state
variable x(t+1) of the world at time t+1 as a function of the state
variable x(t) of the world at time t. Which brings us to {\it Markov
chains}.

A Markov chain is a Markov process whose transition matrix is the same
at all times t (see Section 3.1 for the definition of a Markov transition matrix.)
Call this unchanging matrix G the generator of the process.  Given G
plus the state at time 0, we can calculate p(x(1)), p(x(2)) .. etc., and
also p(x1,x2) and p(x2,x3) .. etc.  It follows from the Markov property
that we can calculate the whole joint distribution (this is a well-known
theorem.)

A generator, in contrast to an initial state, is the {\it law} of a process.
It's important to realize that not every Markov process has such a law.
Given the Markov property alone, the matrix T of conditional
probabilities between successive states can be an arbitrary function of
time, in which case the distinction between law and initial condition,
between ``essence" and ``accident", is artificial at best.

What happens to the generator if we reverse time?  Except in very
special cases, it does not exist.  The transition matrices of a
time-reversed Markov chain are almost always time-dependent.  This makes
the chain property a good indicator of time direction---score one more
point against Hume!  It's far from being a definitive indicator, though,
since this reversed time-dependence is always lawful, and the reversed
chain can always be embedded in a larger process with the chain
property.

We have not yet mentioned systems; the concept of generator applies only
to the process itself.  However, it follows quickly from our results in
Section 3.5 that we can always represent a Markov chain by a link system
whose components, except for the first one, look and behave just like G.
More exactly:

\noindent
{\bf Causal link chain theorem}:  Given a Markov chain C with generator G and
initial vector V, there is a link system whose first component is V and
whose successive components all have the matrix (1/n)G, where n is the
{\bf dimension} of V etc., i.e. the number of values in the range of the state
variables.

Proof:

Let x be the initial chain variable, and x',y be the variables (indices)
of G. Then we must show that the joint distribution on x, y that results
from assigning x', the conditioning index in G, to x, is the same as
that resulting from linking the joint distribution in x', y having the
joint distribution matrix (1/n)G to the independent variable x with the
link (x=x').

Recall that the sum of every column of a transition matrix is 1, so
(1/n)G is a joint distribution matrix in which the variable x' is white.
Thus, by the white connection theorem, linking x and x' does not change
the distribution on x, and so multiplies the i'th column of (1/n)G by
p(x)/p(x=y) = np(x).  But this is just what you get when you multiply
the initial vector by the transition matrix G.

\noindent
QED.

\noindent
{\bf Causal link chain}:  The link system derived from a Markov chain as
above.  When we don't need to carefully distinguish system from process,
we'll usually call such a system a Markov chain too.

\noindent
{\bf Causal transformation}:  A transformation T in which the probability
distribution on the row index is white, i.e. all columns have the same
sum.  The transformations in a causal link chain are of course causal.
A transition matrix as ordinarily defined is a causal transformation
for which all elements are non-negative.

\noindent
{\bf Theorem}:  The product TU of causal transformations T and U is causal.

\noindent
{\bf Theorem}:  The link states of a causal link chain are of the form 
$\vert$v$><$w$\vert$,
where w is white.

We have now mapped Markov chains onto link chains in such a way that
their generators turn into causal transformations with the same
matrices, except for a normalizing factor n. It was important to keep
track of such {\it normalizers} in proving our basic theorems, but in the
future it will make life easier to ignore them until it comes time to
compute actual probabilities.  This is almost always possible because,
if we know any multiple of a probability distribution on x, we can
divide by its sum to get the probabilities themselves.  When it comes to
dealing with negative probabilities, there may be no sensible normalizer
at all.  For instance, it can happen that even though all the case
counts we need are well defined, the total case count is 0.

The passage of time in a Markov chain can be represented by a
time-dependent transition matrix T(t), where T(t+1) = T(t)G.  By the
product theorem in Section 3.5, this same ``transition law" (modulo a
normalizer) is true in a causal link chain.  The two concepts differ
radically, however, when it comes to the meaning of the word `state'.
In a Markov chain, the states are what we have called the state vectors
of the variables, which directly represent their probability
distributions.  These state vectors are indeed still well-defined in
link theory, and have the same meaning.  However, the word `state'
itself has been given a very new meaning, and now refers to the density
matrix of a link.  Since the state vector is the diagonal of the state
matrix, probabilities are gotten by the rule p(P) = tr(PS), as we saw in
the last section.

In Markov chains, the law governing change of state, or as we'll call
it, the {\it dynamical rule}, is V' = T(V), where T is the time-dependent
transition matrix, V the initial state, and V' the final state.  Things
are very different when we come to link states.

\noindent
{\bf Link dynamical rule}.  Let T be a causal transformation.  If the matrix
of T has an inverse, then S' = TST$^{-1}$, where S is the state before T, S'
the state after.  If T does not have an inverse, then the most that can
be said is that TS' = ST

Proof.  Let x' and y be the variables of T, where the links to T are
x=x' and y=y' for external variables x and y'.  Now consider the join of
the components other than T, and abstract from this the sub-process with
variables x and y'.  Cut the link x=x'.  By the product theorem, the
matrix S in the unlinked variables x and x' is UT.  Similarly, the
matrix S' in the unlinked variables y and y' is TU.  We now have S = UT
and S' = TU.  Thus S'T = TUT = TS.  If T has an invertible matrix, then
we can multiply on the right by T$^{-1}$ giving the stronger rule S' =
TST$^{-1}$.

The classical dynamical rule is the offspring of Newtonian determinism,
and is based on a certain concept of explanation, which equates our
understanding of a phenomenon with our ability to reliably predict it.
With the return of chance, reliability becomes a matter of degree.  The
best laws are now those that give us the best estimates of probability.
Nevertheless, the aim of classical explanation remains the same, namely
reducing information about the future to information about the past.
Thus the form of an explanatory law is that of a function that yields a
probability distribution on the future as a function of a probability
distribution on the present.

Notice that all this changes with the link dynamical rule.  S' is only a
function of S if T is invertible.  Otherwise, the dynamical rule only
specifies a certain relationship between future and past, in which the
two enter symmetrically (note that if T is invertible, we can also write
S= TS'T$^{-1}$.)  With this shift come new questions about the very meaning
of dynamical explanation.  The traditional scientist asks {\it how} the past
dominates the future, but never {\it whether}.  But link dynamics reveals many
other kinds of relational structure besides domination.  We are
beginning to see hints of how link theory could enable us to remove our
``causal colored glasses".

\noindent
{\bf Differential Markov chain}:  One whose generator makes a very small
change in the state.

It greatly simplifies many calculations to write the generator of a
differential Markov chain in the form G = 1+Hdt, where 1 is the identity
matrix.  Ignoring second-order effects, we then have G$^2$ = 1+2Hdt etc.,
so we can calculate T(t) by integration rather than by repeated matrix
multiplication.  In differential Markov chains, it's customary to speak
of H rather than G as the generator of the process.  Let $\psi$(t) be the
state vector of the chain at t, and define d$\psi$(t) as
$\psi$(t+dt)-$\psi$(t); then
for any t we have d$\psi$ = G($\psi$) - $\psi$ = H($\psi$)dt.  
Many people have remarked on
the resemblance between the equation d$\psi$ = H($\psi$)dt
and Schrodinger's equation.  We'll now see that this resemblance is no accident.

First, note that 1+Hdt is always invertible, it's inverse being 1-Hdt
(we are ignoring second-order terms, so (1+Hdt)(1-Hdt) = 1-H$^2$dt$^2$ = 1).
Treating 1-Hdt as a generator would give us our original Markov chain
reversed in time!

But doesn't this clash with our earlier observation that the
time-reversal of a Markov chain isn't a Markov chain?  No, because
closer inspection reveals that 1-Hdt always contains negative
``probabilities".  Of course if we allow negative probabilities, then the
answer is yes, and we have a useful tool for retrodicting state vectors.
However, be careful here.  Though the state evolution is reversed by the
inverse process, the forward process itself, which includes the joint
distribution on all variables, is not, since the inverse process always
has negative joint probabilities.

\noindent
{\bf Inverse Markov chain}:  A formal chain with a generator that is the
inverse of a causal transformation.

If we encounter a time-reversed Markov chain, there are two things we
can do in order to preserve the chain property:  We can turn ourselves
around in time, in effect putting the boundary condition on the future,
or we can allow negative probabilities and work with the inverse chain,
which means giving up studying the transitions and contenting ourselves
with state vector dynamics.  Suppose, though, that we encountered a
chain C made up of two chains C1 and C2, the one forward and the other
reversed.  Then it would seem that we only have the second option.
However, link theory gives us a third, which is to put a separate
condition on both the past and future of C, where the initial condition
is white for C1, and the final condition is white for C2. This is 
illustrated in Fig. 1.

The ``generator" G = G1\&G2 of C is of course not a causal transformation,
so C is not a Markov chain.  But G persists throughout the time of the
process and thus should qualify as a law.  What is essentially new is
that in order to define the process as satisfying this law we must now
be given two boundary conditions, one initial and one final, and that
without this final condition, the process has no law!  This observation
leads us to a basic generalization of the concept of Markov chain:

\noindent
{\bf Chain}:  A link system given by a repeated transformation G called a
generator, an initial component V and a final component W. This is
illustrated in Fig. 2.

As we'll now see, quantum systems result from opening up our formalism
to allow for both negative probabilities, and chains with double
boundary conditions.

We've been focusing on transformations; let's now turn to states.  As we
saw above, the link states of a causal chain are of the form
$\vert$v$><$w$\vert$,
where w is white.  We know from quantum physics that quantum density
matrices are self-adjoint, which means that pure quantum states are of
the form $\vert$v$><$v$\vert$.  What kind of link chains have quantum states?

Suppose the initial link state of a quantum chain is $\vert$v$><$v$\vert$.  By the
dynamical rule, the second link state will then be 
G$\vert$v$><$v$\vert$G$^{-1}$.  For the
second state to be quantum, i.e. to be of the form $\vert$v'$><$v'$\vert$, we must
have $<$v$\vert$G$^{-1}$ = (G$\vert$v$>$)$^{\dagger}$ = $<$v$\vert$G$^{\dagger}$.  
(Recall that T$^{\dagger}$ is the adjoint of T,
which for real matrices is just T with rows and columns reversed.)  If
this equality is to be true for any initial state vector $\vert$v$>$, i.e. if we
are to be able to separate the law of the process from its boundary
conditions, then we must have G$^{-1}$ = G$^{\dagger}$.  But a transformation whose
inverse is its adjoint is a unitary transformation.

Thus we see that the generator of a quantum chain is unitary.  One
question remains; Given the initial vector $\vert$v$>$, what must be the final
state vector $<$w$\vert$?  Let T be the product GGGG... of all the G's.  Then
$\vert$w$><$w$\vert$= T$\vert$v$><$v$\vert$T$^{-1}$, 
so $<$w$\vert$ = \break
$<$v$\vert$ T$^{-1}$.  We can write T$^{-1}$ as
G$^{-1}$G$^{-1}$G$^{-1}$G$^{-1}$.... which leads to a way of diagramming quantum chains that
is very useful in analyzing quantum measurements, as we'll see in the
next section.  First an important definition:

\noindent
{\bf Sharp vector, state}:  A vector is called {\it sharp} if it contains a single
non-zero entry.  A quantum state matrix S is also called {\it sharp} if it
contains a single non-zero entry.

Since S is self-adjoint, its non-zero entry must be in its diagonal,
which means it is a pure state whose vector components are also sharp.
This leads to the diagram of a quantum system which is
prepared in a sharp state given in Fig. 3.

As with a differential Markov chain, we can write the equation of a
quantum chain in the form d$\psi$ = H($\psi$)dt, where $\psi$ is the state vector.  But
where is the i? you ask.  We still haven't arrived at physics yet; what
we have been studying is a more general mathematical structure known as
{\it real} quantum mechanics.  As mentioned, Mackey showed how to get
{\it complex} quantum mechanics as a special case of real quantum mechanics by
introducing a (real) linear operator, which we'll call i, that commutes
with every other operator.  We'll take a slightly different route.

Instead of starting out with i as an operator, we'll start with a matrix
representation of the complex numbers, where a+ib is represented as a
2x2 matrix C$_{jk}$, with C$_{11}$ =C$_{22}$ = a, and C$_{12}$ 
= -C$_{21}$ = b. One can quickly
verify that these matrices behave like complex numbers under addition
and multiplication.  A complex ``probability measure" is then simply an
additive measure on a Boolean algebra having these matrices as its
values.  Let x be a variable on which there is such a complex
probability distribution c(x).  We can map this onto a real distribution
on three variables x, j and k such that for any x, p(j,k) is the j,k'th
entry of c(x).  With this mapping, a complex probability acquires a
definite meaning in terms of real probabilities:  it is a joint
distribution on two binary variables j and k satisfying p(j=1 \& k=1) =
p(j=2 \& k=2) and p(j=2 \& k=1) = - p(j=1 \& k=2).  A complex variable x is
then a process in x, j and k such that, for any m, p(j,k$\vert$x=m) is a
complex probability.

Suppose E and E' are independent events with complex probabilities c and
c'.  What, then, is the probability of E\&E'?  We would of course like it
to be cc'.  However, if c and c' are independent real 2x2 matrices, i.e.
independent joint distributions on i,j and i',j', then combining E and
E' produces a distribution p(j,k,j',k') = p(j,k)p(j',k') which does not
even make sense as a complex probability.  This presents us with a
choice:  We must either give up trying to reduce complex probabilities
to real probabilities, or else give up our old notion of independence,
putting the complex product rule c(E\&E') = cc' in its place.  Since
quantum mechanics uses this complex product rule in defining the inner
product of two quantum objects, we'll take the second course, and ask
what it means when translated into real probability theory.

To multiply c and c' means to {\it contract} k and j', i.e., 
to link k to j'
and then ignore k, or equivalently, since cc' = c'c, to contract j and
k'.  Thus to combine a complex process W with an independent complex
process W' into a complex process W\&W' we must always link a binary
variable of one with a binary variable of the other.  In quantum
mechanics, an n-dimensional unitary transformation T can be thought of
as operation on an object which is an entangled composite of an
n-dimensional real object R and a two-dimensional real object J, where
the entanglements satisfies the skew-symmetry rules above.  The 2x2
pieces of T can be thought of as operators on J. Suppose T' operates on
an independent R'\&J'.  To bring T and T' into the same universe of
discourse, we must link j and j', thus multiplying the transformations
on J times those on J'.  But it only makes sense to multiply
transformations if they operate on the same object, so we conclude that
J=J'!

In other words, there is a particular two-state particle that belongs to
every quantum object; this is what we referred to earlier as the Janus
particle.  That name was based on a conjecture relating its states to
the equivalence of the Heisenberg and Schrodinger representations, so
for present purposes lets' give it another name:

\noindent
{\bf Phase particle}:  The two-state particle that is common to all quantum
objects when we interpret complex amplitudes as 2x2 real matrices.

As is well known, only the relative phase of quantum amplitude can be
observed, the absolute phase being unobservable.  This is equivalent to
saying that the state of J is unobservable.  Note that this means that
``pure" quantum states, regarded as real states, are always mixtures with
two-dimensional degeneracy.  An interesting way to construct complex
quantum mechanics is to start with real objects Ri, add J to each of
them, and then confine ourselves of those states of Ri\&J in which either
J is independent of Ri (in the real sense), or is entangled in such a
way that changing the state of J does not affect the probabilities in
measurements of Ri..

\noindent
{\bf Complex quantum mechanics}:  Real quantum mechanics where everything
intersects in a single quantum bit whose state doesn't matter.

\noindent
{\bf Janus particle}:  A hypothetical two-state particle belonging to all
quantum objects, among whose symmetries are the reversal of the
Heisenberg and Schrodinger representations.  Let us here go on record
with our conjecture that the Janus particle is the phase particle, and
that resulting ``subject-object symmetry" is the conjugation symmetry of
quantum mechanics.

\subsection{The Quantum Measurement Problem and its Solution}

{\it The measurement problem in link theory, the paradoxes of negative case
counting, standard measurement theory, the projection postulate, the
disturbance rule, the three axioms that define laboratory objects,
functional objects, experiments, confined quantum objects, quantum time
vs. laboratory time, prepared quantum chains, definite present and open
future, the projection postulate in link theory, the 
collapse of the wavefront, the confined quantum object is a laboratory
object, the white state, time loops revisited, macroscopic matter.}

The Born rule tells us that, for any proposition P and quantum variable
x, the probability of a measurement of x satisfying P is tr(PS), where S
is the link state of x. The dynamical rule tells how the state S changes
with the passage of time.  So doesn't that wrap it up?  What more is
there to be said about quantum observation?  In the standard Hilbert
space approach, there is the problem of how to construe our ``classical"
measuring instruments as quantum objects, which led von Neumann to his
famous construction of the movable boundary between object and observer.
But this problem doesn't arise in link theory, since quantum and
classical can perfectly well coexist in the same joint distribution.
Indeed, as we have seen, the very distinction between classical and
quantum rests in part on arbitrary decisions about how to analyze a
process.  So what, then, is left of the infamous quantum measurement
problem?  Have we in fact solved it?

Not quite.  In fact, the measurement problem takes a particularly acute
form in link theory, as we'll now see.  Fig. 4a shows a single complete
measurement of a variable x in a pure state $\vert$v$><$v$\vert$; we simply draw an
arrow from x to a recorder, linking x to an input variable of the
recorder which is assumed to be white.  So far so good; there's no
measurement problem yet.

If we wait until time t(y) to measure (Fig. 4b), then the state will
have been transformed into TST$^{\dagger}$, and an arrow from y to the recorder
will again give results in agreement with standard theory.  Still no
measurement problem.

But now let's consider two complete measurements, one before and one
after transformation T (Fig. 4c).  Each of these measurements will give
the same probabilities that it did when it was the sole measurement.
However, that's not what the standard theory predicts.  What went wrong?

Here's one thing that went wrong.  Consider a {\it selection} from among all
the records of just those records for which the first measurement had
the value k. Within this subset, every event in x will have probability
1 or 0, so x is independent of all other events, and v, which we'll
assume is not sharp, is replaced by a sharp vector v'.  The result will
be in most cases that the state of y is no longer quantum.  In other
words, measuring the object throws it out of a quantum state.

But we're in worse trouble than that.  In the system of Fig. 4c, it may
well happen that some of the joint probabilities on x and y are
negative, which means our recorder has recorded ``negative" cases!  So
what's wrong with recording negative cases? you may ask.  Remember,
recorded cases are not just alternatives, they are {\it instances}; they are
actual things, marks on a tape, whatnot.  Suppose your closet contains
ten positive white shirts and ten negative green shirts.  If you reach
into your closet for a shirt, you will come out empty handed, since
there are 0 shirts there.  On the other hand, if you reach in for a
white shirt, you'll have ten to choose from.  That's quantum logic.

To see the problem of repeated measurement more sharply, consider a
third measurement on z which comes after a second transformation T$^{\dagger}$
(remember, T is unitary, so T$^{\dagger}$ = T$^{-1}$); this is shown in fig. 5.

As before, we'll run a long series of recorded experiments.  Since TT$^{\dagger}$ =
1, x and z are perfectly correlated in this process.  Since the recorder
is white and therefore doesn't effect the probabilities of x, y and z,
this perfect correlation will show up in the total set of records, i.e.
x and z are equal in every record.  But now consider the subset S$_k$ of
all records for which y=k.  Since y separates x and z, we know that x
and z are conditionally independent given y, which means that they are
independent within S$_k$.  Thus the records in S$_k$ will in most cases show x
and z as having different values!  S$_k$ is like your white shirts in the
closet---it's a non-null subset of the null set.  But real shirts and
real records just don't behave like that.  Obviously we've got to take
another tack.

Up to now in this paper we have managed to avoid the logical anomalies
that result from negative case counting by carefully steering away from
them, the trick here being to think only about the arithmetic of ratios.
But this won't work forever.  Sooner or later we have to face the
paradox of the white shirts.  This is the paradox of the two-slit
experiment seen straight on.  With both slits open, the case counts
cancel at the null in the ``wave interference pattern" where the detector
lies, and no electrons are detected.  But with one slit closed, the
``white shirt" electrons alone are selected, and they make it through.
We'll see that the solution of the measurement problem involves
transforming this seeming paradox into a mathematical account of what it
means for the future to be {\it open}.

Since the predictions of the standard theory of quantum measurement are
in close agreement with experiment, we must represent measurement by
link systems that make these same predictions.  To see just what is
required, here is a brief summary of the standard theory, which is based
on an assumption that von Neumann called the {\it projection postulate}.

\noindent
{\bf Projection postulate}.  Given a state S and a proposition (projection) P,
there is an ideal measurement M to test for the truth of P that leaves
those things that pass the test in the state PSP/p, and those things
that fail in the state P'SP'/(1-p), where p = tr(PS) and P' is the
negation 1-P of P.

It follows directly that M leaves the object in the state PSP+P'SP'.
Let Q be any observable with eigenvalues q$_1$, q$_2$, ..q$_n$, 
and let P$_1$,
P$_2$..P$_n$ be the ``projectors" of these eigenvalues, i.e., the (ortho)
projections whose ranges are the eigenspaces of q$_1$, q$_2$..q$_n$.  It is easy
to show from the projection postulate that we can measure Q by
performing a series of tests for, P$_1$,P$_2$..P$_n$.  The object will only pass
one of these tests, P$_i$, which tells us that Q has the value q$_i$.  The
ensemble of objects thus tested will be in the mixture
$\Sigma_i$P$_i$SP$_i$, which
gives us the general rule for the minimum ``disturbance" of a state by a
measurement.

\noindent
{\bf Disturbance rule}:  Given an object in state S, the least disturbing
measurement we can make of quantity Q is one which leaves the object in
the state $\Sigma_i$P$_i$SP$_i$, where the P$_i$ are the projectors of Q.

The disturbance rule gives us a way to prepare a beam of objects in any
state S, which is to start with a beam in any state and measure it in a
way that yields the components of S, and then select out a subset of
objects from each component to give the proper weight of that component
in S.

Practically speaking, the standard theory of measurement works very
well, at least for simple systems.  But the projection postulate has
caused a lot of trouble for people concerned with quantum foundations,
since, by ``collapsing the wavefront", it messes up what is otherwise a
beautifully clean and simple unitary dynamics.  Since link theory has no
wavefront to collapse, it ought to be able to do better, and indeed it
does; in fact, we count its success in this regard as its major
accomplishment to date.  We'll see in a while what the projection
postulate means in terms of linking.  But first we need a description in
our current language of the kinds of things that go on in a physics
laboratory.

The first thing to be said about our idealized and simplified laboratory
is that there is a clock on the wall.  Someone, or something, is always
watching that clock and logging every event, so that at the end of every
experiment there is a diary of the experimental procedures and the
outcomes of measurement..  The second thing to be said is that these
experimental procedures can be repeated as often as necessary to obtain
meaningful relative frequencies.  Let's now characterize the {\it
objects} in our lab.

\noindent
{\bf Laboratory object}:  A process K in which we distinguish two kinds of
{\it external} variables, called {\it input} and {\it output} variables.  Each such
variable exists for only one clock tick; for instance x could be the
keyboard input to a computer at time t; we'll write the time at which x
occurs as t(x).  Input variables are those that we control, while output
variables are those that we measure, or use to control inputs.  K may
have other variables besides its inputs and outputs, and it need not
have inputs at all, though it must always have at least one output.

When we speak simply of {\it the input} to K, we'll mean all of its input
variables together, or more exactly, the join of all these variables.
Similarly for {\it the output} from K. The input (output) {\it before} or
{\it after} time
t$_1$ is the join of input (output) variables x$_i$ such that t(x$_i$)
$<$ t$_1$, or t(x$_i$) $>$ t$_1$.

We'll assume that laboratory objects satisfy three axioms.

\noindent
{\bf 1) Controllability}:  The input is white.

\noindent
{\bf 2) Causality}:  Given the input before t, the output at or before t is
independent of the input at or after t (no influences backwards in
time.)

\noindent
{\bf 3) Observability}:  The output never has negative probabilities.

These three axioms broadly capture our common-sense notion of processes
that we can observe and manipulate.  Notice that they have nothing
whatsoever to do with classical mechanics.  Bohr characterized the
laboratory as the place where Newtonian mechanics reigns, but this was a
mistake.  Newton's laws neither imply nor are implied by the above
axioms; indeed they belong to a wholly different universe of discourse
(that's one reason why the young Russell, who admired physics more than
common sense, was so eager to jettison causality.)

\noindent
{\bf Functional object}:  One in which the output is a function of the input.
The requirement of whiteness on the input imposes the condition on the
joint distribution that all probabilities are either 0 or 1/n, where n
is dimension (total number of cases.)  It's convenient to ignore
normalization and represent such a joint distribution by the bit array
which is the characteristic function of the object's function regarded
as a relation.

\noindent
{\bf Experiment}:  A link system whose components are laboratory objects, one
of which is a functional object known as a {\it recorder}.  The output of the
recorder is a variable at the end of the process whose value is a record
of all inputs and outputs to the object on which the experiment is
performed.  It is assumed that every input variable in an experiment is
linked to a prior output variable, and from this it follows that the
process represented by the experiment is itself a laboratory object.

Let's now turn to the construction of quantum objects.  We'll show these
as enclosed in a dotted box, as in Fig. 4 and Fig. 5.  Our aim now is to
design dotted boxes that enclose their quantum weirdness, so-to-speak,
i.e., boxes whose in-going and outgoing arrows satisfy our definition of
a laboratory object.  Such confined quantum objects are, properly
speaking, the true objects of scrutiny in a quantum physics lab.

Notice that the dotted boxes in Fig. 4 are confined quantum objects;
they satisfy 1) and 2) because they have no inputs, and 3) because
their probabilities satisfy Born's rule and are therefore non-negative.
The dotted boxes in Fig. 5 are not confined objects, however, since
their outputs, i.e. the joint distributions on x, y and z, do have
negative probabilities.  Our challenge now is to redraw Fig. 5 as a
confined object that gives the right numbers.

There's one thing doesn't look right about the box in Fig. 5, 
which is that to
prepare it in state $\vert$v$><$v$\vert$ requires placing a boundary condition both
before time t(x) and after time t(y).  If we think of these conditions
as inputs from the lab, the second condition violates 2), the causality
principle.  This is assuming that the time sequences inside and outside
the box are the same.  But there is another possibility, and a more
likely one, which is that ``time" in the quantum domain is somewhat
autonomous from laboratory time, or to put it more abstractly, that the
chains of quantum separability that we compared to Markov chains are in
themselves not temporal sequences at all.  Under this assumption, it is
reasonable to define a {\it prepared} quantum chain as one whose beginning and
end are both at the initial laboratory time of the experiment (Fig 6a).

This construction looks more natural when we draw it with T$^{\dagger}$ factored
out of w (Fig. 6b).  Reversing rows and columns of T$^{\dagger}$ turns 6b into the
equivalent form 6c.  Let's then take this as our picture of a chain
prepared in a pure state; notice that the states S and S' are identical.
More generally:

\noindent
{\bf Prepared quantum chain}:  A parallel pair of identical unitary chains
whose initial vectors are the same and whose final variables are linked.
(Fig. 7).

Suppose that when we measure x, the future measurement of y is
indeterminate, in the sense that all outcomes are possible..  This
indeed is our experience of time, which we express by saying that there
does not exist at time t(x) a record of the outcome of the measurement at
t(y).  Thus it makes no sense to speak at t(x) of the subset of records
for which y has a certain value.  But this implies that x and x' are
linked at time t(x) by the matrix TT$^{\dagger}$ = 1, which means that recording
the value of x at t(x) also records the value of x' at t(x).  Since we
must sum over all values of y to evaluate probabilities at t(x), this
avoids the paradox of Fig. 5.

Now what happens when we measure y?  In our earlier example, this
measurement unlinked x and z, making them in fact independent.
However, in the present case, x' is already on record as having the same
value as x, so it of course cannot become independent.  How do we show
this in our diagram?  Simply by drawing a line between x and x' (Fig.
8a).  Recall that a recording device is a laboratory object having
(white) inputs whose output at the final time is a 1-1 function of the
history of its inputs.

Before y is measured, this line merely duplicates TT$^{\dagger}$, so drawing it
would be redundant.  But after y has acquired a definite value, this
line in effect remembers the measurement of x', thus holding the fort
against the disruption that would otherwise be produced by fixing y.
Therefore there will be no records at all where x and x' are unequal,
and the paradox of a non-null subset of the null set is avoided.

Choosing to measure x draws a line between x and x' as well as a line
from x to the recorder.  This is the link theory version of the
projection postulate.

\noindent
{\bf Probe x}:  Link x to x' and link x to the recorder (strictly speaking,
this is a complete probe; we'll come to incomplete probes later.)

\noindent
{\bf Projection postulate for a complete measurement}:  To make a complete
measurement of the state of a prepared quantum chain at time t(x),
insert a link between x and x' and measure the linked variable x, i.e.
{\it probe} x.

The choice of whether or not to probe x can be thought of as a
double-pole switch.  Just what do we mean by the term `switch'?  By a
{\it single-pole switch} we mean a three-variable component Sw[s,x,y] where s
is a binary variable with values ON and OFF, Sw[ON,x,y] is the identity
matrix, and Sw[OFF,x,y] is the white matrix in which every element
equals 1/n, n being the dimension of x and y. Notice that for unlinked x
and y, s is white, since the sum of matrix elements in the ON condition
is the same as that in the OFF condition.  By a {\it double-pole
switch} we mean a pair of single-pole switches whose s's are linked.

The crucial question now arises whether this switch input satisfies
axioms 1) and 2), i.e. whether it is white and causal.  If so, then the
dotted box qualifies as a laboratory object.  To say that the switch is
white means that the probability P of the process is unchanged by
drawing the line.  By the summation theorem, P is the sum of the joint
distribution on x and x'.  Drawing the line eliminates all off-diagonal
terms in this sum.  But since TT$^{\dagger}$ = 1, these off-diagonal terms are all
0 anyway, so the line doesn't affect P. This takes care of 1); we'll
come to 2) shortly.

Looking at Fig. 8b, we see that closing the switch breaks the diagram
into two parts separated by x. The second part reduces to a circle, as
far as x is concerned, so its presence at x is white.  Thus the state S
of the link from v to x is unchanged by the measurement of x; it is
still $\vert$v$><$v$\vert$, and the probability distribution on x is still the
diagonal of S, in accordance with Born's rule.  But what is state S$_1$?
Clearly it has the same diagonal as S. But, since TT$^{\dagger}$ = 1, its
off-diagonal elements are all 0. S$_1$ is then the mixed state having sharp
components weighted by the probability distribution on x. Thus we see
that this change of state from S to S1 satisfies the standard theory of
measurement, and is indeed nothing else but our old friend the collapse
of the wavefront!  Of course there is now no longer a wavefront to
collapse, so we'll have to find another name for it.  Whatever we call
it, though, it's the crucial event that links the micro to the macro,
and it is brought about by the {\it passive} act of recording the state of x.
Our probe does not disturb what it sees; its effect comes from seeing
alone.  

That seeing alone is sufficient to collapse the wave function
was a point made by HPN in the context of the double-slit
experiment performed with low-velocity cannon balls, armor plate and
mechanical detectors \cite{Noyes80}. This point needs a little further
discussion here in anticipation of both more abstract {\it and}
more physical discussions of the measurement problem which are in
preparation. ``Seeing'' the low-velocity cannon ball go through one
slit or the other requires illumination, and reflection of at least one
photon from the cannon ball itself, together with reflection of enough
light from the two slits to distinguish which it went through.
Hence there is actual energy transfer from the ball to the observer.
This energy transfer has to be recorded before this particular case can
be used to enter as a record in the compilation of single slit cases
in contrast with double slit cases. The act is ``passive'' but not
devoid of physical content and (pace von Neumann and Wigner) does {\it
not} require a {\it conscious} observer to ``collapse the wave
function'', as we hope the analysis made in terms of the abstract
theory presented above has made clear. More generally, making the
``quantum system'' into a ``laboratory object'' requires conscious
experimenters to set up the apparatus and to interpret the results
after they are recorded, but this in no way distinguishes it from the
protocols of classical experimentation where no question of quantum
effects arises. When such processes are studied in enough detail
to observe the progressive (and predicted) ``randomization'' of
mesoscopic quantum states --- as has recently been done \cite{Haroche98}
with two separating Rydberg atoms in $n_{Bohr} = 50$ states ---
the quantum {\it theoretical} treatment must include many more
degrees of freedom than are usually included when discussing
quantum measurement and the projection postulate. We believe that it
might prove instructive to see how much of the analysis of such systems
depends on how they are ``carved up'' between classical and quantum
degrees of freedom and where considerations of experimental accuracy
have to enter.

Closing the ``record" switch does not change the observed probability
distribution on x, and in this sense it does not disturb the
measurement.  However, it does affect the state of y, which would be
TST$^{\dagger}$ with the switch open but is T S$_1$T$^{\dagger}$ 
with the switch closed.  From
the point of view of the laboratory, closing the switch causes a change
in the state of y. Inside the dotted box, though, there is nothing
resembling cause and effect; indeed, nothing even happens inside the
box, since closing the switch merely duplicates an existing link.

Measuring y is just like measuring x, the only difference being that
S$_1$
is mixed (Fig. 8b).  Thus we see that throwing the switch at y doesn't
affect the measurement of x, or the whiteness of the x switch, so the
switch inputs satisfy 2), the causality axiom.  This shows that the
dotted box is a classical object, and the sequence of measured variables
is in fact a Markov process.

So far we've only been considering complete measurements.  An incomplete
measurement of x is represented by a functional box f(x) going from x to
the recorder, where f is not 1-1.  Recall that, in a complete
measurement, connecting x to x' means that by recording x at t(x) we
have also recorded x' at t(x).  However, in the case of a partial
measurement, it's not x but only f(x) that gets recorded at t(x), so
this connection should go from f(x) to f'(x') rather than from x to x'
(Fig. 9).

This arrangement satisfies the standard general rule S'
=$\Sigma_k$P$_k$SP$_k$ for
the change of state due to a partial measurement, where the P$_k$ are the
projectors of the measured observable.

What about preparation?  As remarked, a preparation is always equivalent
to a selection based on a measurement.  But we can also represent a
preparation as an input to an unprepared system, i.e. a system in the
white state (fig. 10a); clearly such an input is always white.

The white state, whose matrix is 1, has a unique role in quantum theory,
since it is unchanged by measurement (i.e., if S=1, then 
$\Sigma_k$P$_k$SP$_k$ = $\Sigma_k$P$_k$ = S).  
The diagram of an unprepared and unmeasured quantum object is
simply a circle (Fig. 10b); we can think of it as representing pure
entropy.  Perhaps this is the only quantum diagram that makes sense in
isolation from the classical world.

Our conclusion that the ``collapse of the wavefront" requires an open
future seems to reintroduce an absolute arrow of time.  But this is an
illusion, since an unprepared system is symmetrical with respect to past
and future, whether it is measured or not.  Time's arrow comes not from
quantum openness but from classical preparation, which we have
conventionally placed at the beginning of the process, but which could
just as well occur at the end, or at both the beginning and end, since
it is equivalent to a selection of records.  Thus our present analysis
is consistent with our previous notion of the ``state throttle" in a time
loop.

The topic of measurement, especially partial measurement, deserves a
much fuller treatment than we have given it here.  However, even the
present abbreviated account calls for a few more remarks concerning
macroscopic matter:

In our analysis above, we have treated quantum and laboratory objects as
essentially different kinds of things.  And yet sooner or later we must
face the fact that laboratory objects are made out of atoms, and try to
render an account of just what this means.  In standard quantum
mechanics, when you assemble quantum objects you always get other
quantum objects, and this indeed does create insoluble problems, since
there's no way to ``bend" quantum-ness into anything else.  The present
approach doesn't have that problem, of course, since it treats
quantum-ness as a mode of analysis rather than an intrinsic property of
things, and there is no reason why the appropriate mode of analysis
should be the same at different scales.  But we can hardly let things go
at that, since there is a huge body of empirical knowledge about the
relationship between micro and macro that needs to be translated into
our present conceptual scheme.

In the last century, the laws of mechanics were thought to form a closed
system that completely governs physical change, all other physical facts
being relegated to the category of ``accidents" of the world state.  How,
in such a scheme, is one to deal with the undoubtedly lawful character
of thermodynamic events like heat flow?  The nineteenth-century answer,
which never quite satisfied anyone, was to impose a secondary layer of
``laws among the accidents", so-to-speak, these being somehow based on
the laws of probability.  With the advent of quantum mechanics this idea
becomes pretty far-fetched, since the equation governing quantum
mechanical change has, except for the factor i, exactly the same form as
the diffusion equation, which is one of the alleged laws among the
accidents.

Let us hope that this classical two-layer concept of law is now well
behind us.  How, then, do we describe the relationship of micro to macro
without it?  Let's start at the level of process.  Remember, in a
process as such there are no states until you decide how to carve it up,
only correlated variables, so it literally makes no sense to ask whether
a process is quantum or causal.  It does, however, make sense to ask
whether or not its joint distribution contains negative probabilities.

Let's pretend for the moment that it doesn't.  What we are then given is
a huge number of micro-variables associated with individual atoms plus a
much smaller number of macro variables associated with regions large
enough to contain many atoms, the latter often closely corresponding to
sums or averages of the micro variables.  The macro-variables can be
directly and repeatedly measured without disturbing them, and they seem
to satisfy pretty well our axioms for laboratory objects, at least for
simple inorganic objects.  We implicitly assume that spatial boundaries
inside the object correspond to separation boundaries in our present
sense---one sometimes \break
 speaks of this assumption as non-locality.
Applying our fundamental theorem makes it possible to define states at
these boundaries, thus turning the macro-behavior of the object into a
composition.  This is essentially how an scientist carves something up
when he wants to make a computer model of it, and there is no doubt that
it works very well.

The reason it works is that macro-behavior has a good deal of autonomy
from what we assume about the behavior of the unmeasured
micro-variables, and for most purposes we can ignore quantum effects
completely, at least under average terrestrial conditions.  It might be
supposed that this is simply a consequence of locality plus the
complexity of the processes the macro-variables are averaging over, and
indeed it would be, if the micro-probabilities were always positive.

However, with negative micro-probabilities the situation is quite
different.  Consider a chain whose generator is periodic, i.e., whose
n'th power is the identity for some n. With only positive probabilities
allowed, such a generator would have to be deterministic.  But this is
not true with negative probabilities, and physicists are well acquainted
with periodic unitary transformations (indeed, any self-adjoint unitary
transformation has period two.)  Now a probabilistic chain of period n
is of course separable at every variable, and would thus satisfy
locality if it were strung through a material object.  However, its
first and last variables are perfectly correlated, just as if they were
directly linked!  From the point of view of common sense, this is
non-locality with a vengeance.

Clearly we need a new assumption that keeps such non-local ``virtual
links" from playing a significant role in macro-behavior.  Physicists
already have a name for what is probably the right assumption:
decoherence.  Since this concept is defined in terms of wave functions,
it needs to be reformulated in the language of probability.  And it
undoubtedly needs to be generalized to deal with matter at low
temperatures, and perhaps with organic matter as well.  Though all this
could become technically complicated, it looks straightforward enough.

There is a much more serious problem which is not solved by decoherence,
however, and that is to explain why negative probabilities remain
confined to the micro-realm.  What prevents them from ``breaking out"
into the everyday world and playing havoc with logic?  We avoided this
problem in the case of quantum measurements by assuming that making a
measurement replaces certain ``virtual links" due to quantum coherence by
actual links.  Something of the same sort must apply to the relationship
between a macro-variable and the quantum average to which it
corresponds.  That is, a macro-variable, by being linked to the world at
large, actualizes the virtual link on the quantum level between an
average variable and its time-reversed dual.  To put it more simply,
macro-regions are self-measuring.  Notice that measuring an average of
many quantum micro-variables extracts only a minute fraction of the
information in a pure joint state of these micro-variables, so these
``macro-measurements" are essentially non-disturbing.

To summarize:  If we assume decoherence plus the actuality of virtual
quantum links from averages to their duals, macro-matter becomes
classical.

But before we break out the champagne to celebrate the final triumph of
logic and common sense, there are a few more questions.  Why doesn't the
time symmetry of the microscopic level carry over to the macro-level, at
least to the extent of showing up as independent macroscopic boundary
constraints on past and future?  To what extent does our reasoning in
terms of states bear on the process itself rather than on how we choose
to carve it up?  Might it be that hanging on to our causal framework,
even for everyday objects, exacts a large price in complexity, thus
blinding us to simple and essential forms in nature?  And finally, why
should we assume that we can always find ways to keep the irrationality
of negative probability out of our experience?

\subsection{Things to Come}

This has been a long paper, but we fear it has only scratched the surface
of its subject.  We've concentrated here on the basics rather than on
results.  We believe this was the best course, since these basics take
some getting used to.  But in the long run, it's results that win the
prize.  The history of science is littered with the remains of seductive
theories that led nowhere, von Neumann's quantum logic being a recent
case in point.  The first place to look for results in link theory is
obviously physics, about which we've said relatively little, so here, in
conclusion, is a brief prospectus on several new items of physics that
we see on the horizon.

First, some general remarks about space.  So far, we've mostly been
drawing diagrams rather than maps; in the dialogue between composition
and extension, composition has had by far the louder voice.  But the
world doesn't present itself to raw experience as a neat arrangement of
interchangeable parts---first we have to carve it up.  And before we can
carve it up, we have to map it, or ``mark the carcass", as we put it
earlier.  The places where we mark are the places that separate the
information in the extended structure, the boundaries through which
correlation flows, and which, when rendered ``immobile" by fixing the
values of their variables, break the whole into independently variable
parts.  The question naturally arises as to how the structure of the
probabilistic separating relation p(A\&B\&C)p(B) = p(A\&B)p(B\&C) is related
to more familiar geometric structures like metrics and topologies and
linear orders.

Consider, for instance, a Markov process.  We always start out by giving
its variables in a certain order, but is this order contained in the
probabilities or is it arbitrary?  We saw that the arrow is arbitrary,
so we are really talking about the between-ness relation.  Let's define
B(x,y,z), read ``y is between x and z", to mean p(x,y,z)p(y) =
p(x,y)p(y,z).  We know that separation is equivalent to the Markov
property, which holds for any x,y and z in the given order.  Thus the
question is whether B(x,y,z) always excludes B(y,x,z) and B(x,z,y);
let's call this the order property.  If a process has the order
property, then the linear order of its variables is indeed intrinsic to
that process itself.

There are some obvious instances of Markov processes in which the order
property fails.  For instance, a sequence of independent variables is a
Markov process, but since any one of the variables separates any two
others, the variables can be rearranged in any order; intrinsically,
they are a mere heap.  At the opposite extreme is the case where the
variables are perfectly correlated, i.e., where the transition matrices
are permutation matrices.  If you fix any variable in such a process,
you fix all the others too.  But, as we saw in Section 1, a fixed
variable is independent of all others, so, again, any variable separates
any other two.  It may seem odd to think of Laplace's perfect Newtonian
clockwork universe as a mere disordered heap of states, but there it is!

The true opposite of the independent variable case is the differential
process given by a system where G = 1 + Hdt.  If H is not 0 and there
are no negative transition probabilities, such a process is always
ordered.  Hdt doesn't have to be truly infinitesimal to guarantee this,
only rather small.  If G is constant in time, the probabilities not only
define an order but a metric, at least up to a scale constant.  The
local order of a continuous process is rather curious.  Since it
approaches the deterministic case in arbitrarily small neighborhoods, it
is, so-to-speak, locally disordered.  That is, it would require
arbitrarily many repetitions of the process to statistically determine
the order of points that are arbitrarily close together; this is rather
suggestive of the time-energy uncertainty principle.

The same kind of reasoning can be applied to higher-dimensional
continuous processes, whose probability distributions in general contain
the intrinsic geometry of the space that is used to define them, or at
least its topology.  Keith Bowden\cite{Bowden96} has suggested that if we apply
suitable perturbing random fields to classical fields, we can recover
their spaces from the resulting joint probability distributions.  It
would seem natural in a ``probability space" for the local structure to
be quite different from the global structure, which might have some
bearing on super-string theory.

Now for something completely different..

What's the simplest random process of all?  Surely it's a coin toss.
Let V be a coin tossing process whose HEADS/TAILS variable is x, and
define the velocity v of V as the probability of x = HEADS minus the
probability of x = TAILS (we're not assuming it's a fair coin.)  Notice
that if you step one meter to the right or left every second according
to the outcome of a toss of V, then v is your average velocity to the
right in meters per second.

What's the simplest of all systems with more than one component?  Surely
it's a pair of linked coin tosses.  Call these components V and V', with
binary variables x' and x' and velocities v and v'.  Prior to linking, V
and V' are independent.  Linking x and x' produces the single-variable
binary process U = V\&V'$\vert$x=x', which we can think of as a coin toss too.
What, then, is the velocity u of U in terms of v and v'?

\noindent
{\bf Boost theorem}: u = (v+v')/(1+vv'), i.e., taking the velocity of light be
1, the velocities of linked binary variables satisfy the relativistic
addition law.

\noindent
Proof:  Let p and q be the probabilities of HEADS and TAILS for V, and
similarly let p' and q' for V'.  Then v = p-q and v' = p'-q', and from
the definition of linking one can quickly verify that u =
(pp'-qq')/(pp'+qq').  Thus we must show that (pp'-qq')/(pp'+qq') =
(p-q+p'-q')/[1-(p-q)(p'-q')].  Now in fact these two expressions are not
identical as they stand, but only become identical when we bring in the
additional fact that probabilities add up to one, i.e. p+q = p'+q' = 1.
The easiest way to take these conditions into account is to note that v
= (p-q)/(p+q) and v' = (p'-q')/(p'+q') and substitute these expressions
for v and v' in (v+v')/(1+vv'); the resulting expression then reduces to
(pp'-qq')/(pp'+qq').  QED.

Applied to observer and object, the boost law implies the Lorentz
transformation Thus we find that the concept of linking, which before
led us immediately to the heart of quantum mechanics, has now led us
immediately to the heart of relativity!

There is still a lot of work to be done to relate the above theorem to
the concept of ``probability space" based on separability.  One approach
here may be to interpret ``time lines" as binary Markov chains from which
the LEFT-RIGHT variables are abstracted statistically. 1x1 space-time
would then be the indefinite process that results from linking these
velocity variables in an unspecified collection of such chains.  Notice
the formal resemblance here to our construction of complex amplitudes,
which also resulted from linking an indefinite set of processes via a
binary phase variable.

The question arises whether this resemblance is more than just an
analogy.  Could it be that at some fundamental level, the phase particle
and the ``velocity particle" are one and the same?  Let's briefly
consider where this would lead.  Since in (complex) Minkowski space
boosts are rotations of the complex plane, this identity would make the
relativity of amplitude phase into a generalization of the relativity of
motion.  Even more important for the science of the future is that the
conjugation symmetry of the phase particle would become the symmetry of
v and -v, which is the symmetry that results from reversing object and
observer.  This supports our conjecture that the phase particle is indeed
the Janus particle, and makes more sense of Pauli's mystical vision of i
as the key to uniting physics with psychology, as described in Chapter
2. But of course we are now looking rather far ahead.

\begin{center}

{\bf FIGURE CAPTIONS}

\end{center}

\noindent
Fig. 1. A Markov process in time can be generated either from a vector
given when the process starts and propagate forward in time or
from a vector given when the process ends and propagate backward in
time (see text).

\noindent
Fig. 2. A chain with double boundary conditions. 

\noindent
Fig. 3. A sharp vector linked to a quantum system.

\noindent
Fig. 4. a) A single complete measurement.

b) A single complete measurement following a transition

c) Two complete measurements which cannot reproduce the results of
quantum mechanics (see text).

\noindent
Fig. 5. Illustration of the quantum measurement problem (see text).

\noindent
Fig. 6. a) A prepared quantum chain.

b) A prepared quantum chain redrawn.

c) A prepared quantum chain redrawn again.

\noindent
Fig. 7. General diagram for a prepared quantum chain.

\noindent
Fig. 8. a) A complete quantum measurement.

b) More detailed diagram for a complete quantum measurement.

\noindent
Fig. 9.  Partial quantum measurement.

Figure 10. a) An unprepared quantum system

b) An unprepared and unmeasured quantum object.

\vspace{.5cm}
\begin{figure}[p]
\begin{center}
\leavevmode
\epsfbox{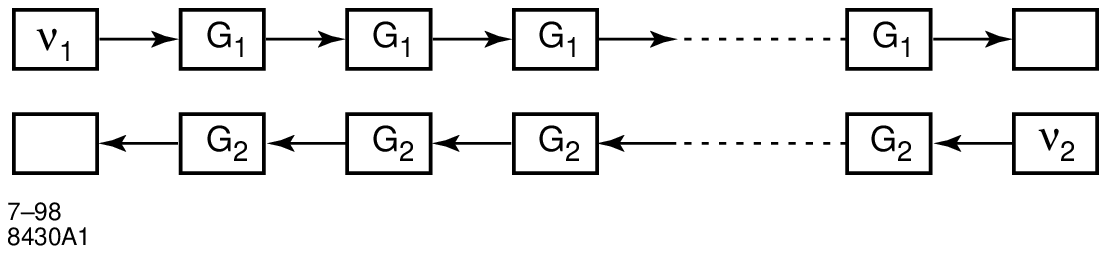}
\end{center}
\label{fig1}
\end{figure}
\vspace{.5cm}
\begin{figure}[p]
\begin{center}
\leavevmode
\epsfbox{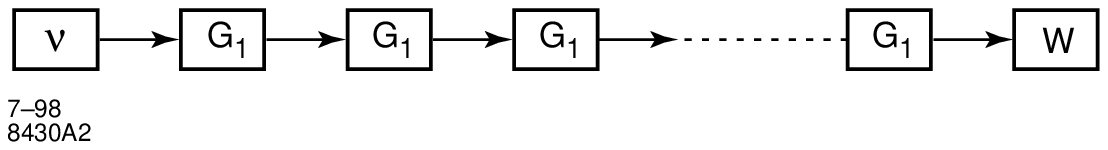}
\end{center}
\label{fig2}
\end{figure}
\vspace{.5cm}
\begin{figure}[p]
\begin{center}
\leavevmode
\epsfbox{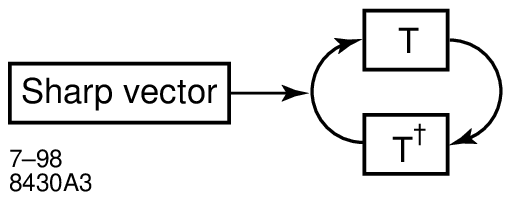}
\end{center}
\label{fig3}
\end{figure}
\vspace{.5cm}
\begin{figure}[p]
\begin{center}
\leavevmode
\epsfbox{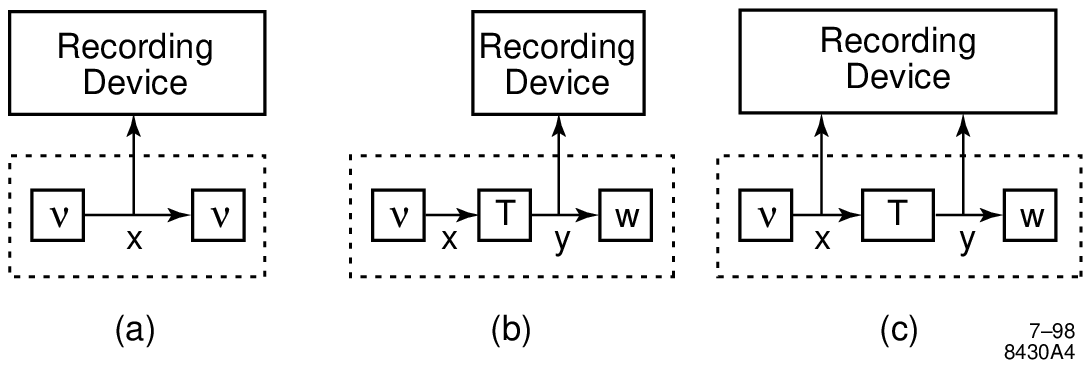}
\end{center}
\label{fig4}
\end{figure}
\vspace{.5cm}
\begin{figure}[p]
\begin{center}
\leavevmode
\epsfbox{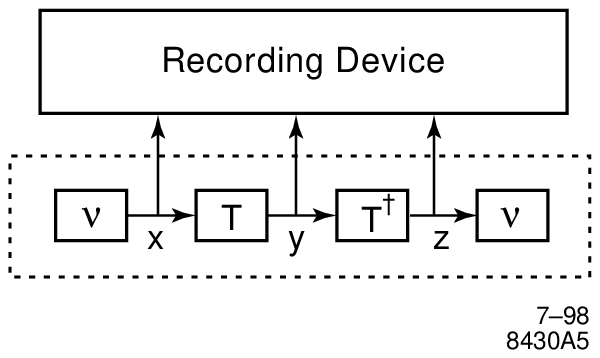}
\end{center}
\label{fig5}
\end{figure}
\vspace{.5cm}
\begin{figure}[p]
\begin{center}

\leavevmode
\epsfbox{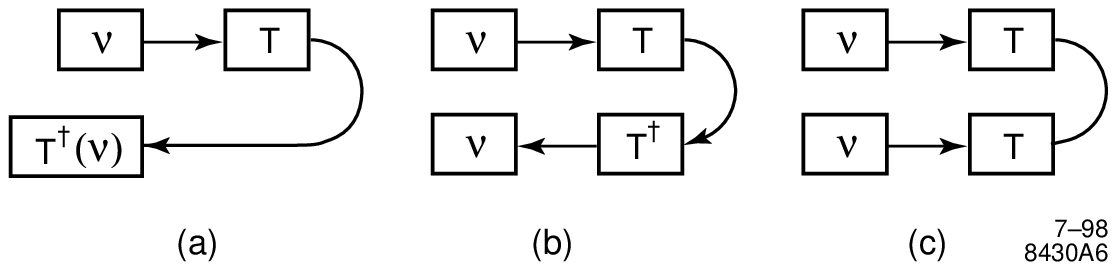}
\end{center}
\label{fig6}
\end{figure}
\vspace{.5cm}
\begin{figure}[p]
\begin{center}
\leavevmode
\epsfbox{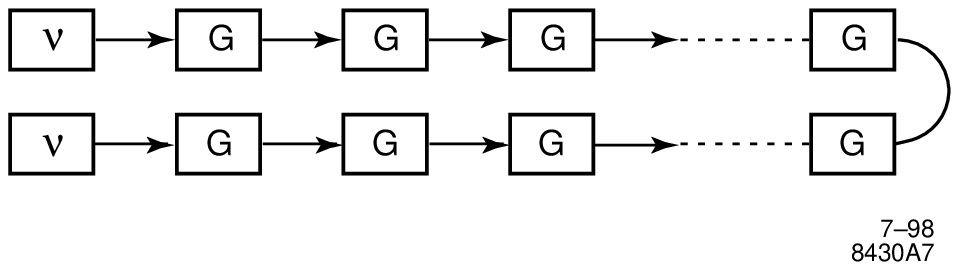}
\end{center}
\label{fig7}
\end{figure}
\vspace{.5cm}
\begin{figure}[p]
\begin{center}
\leavevmode
\epsfbox{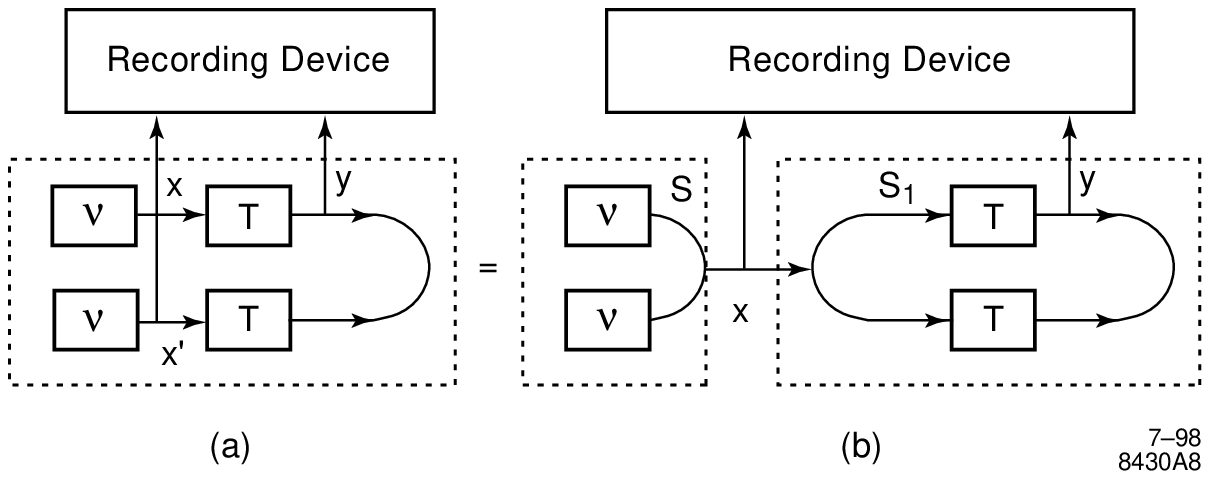}
\end{center}
\label{fig8}
\end{figure}
\vspace{.5cm}
\begin{figure}[p]
\begin{center}
\leavevmode
\epsfbox{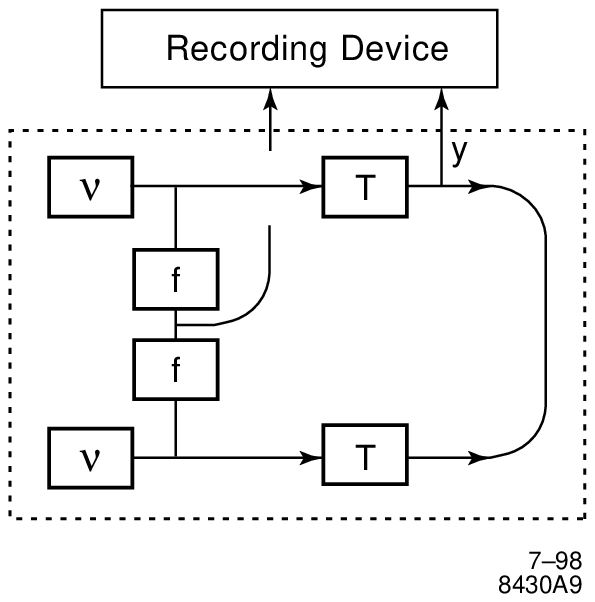}
\end{center}
\label{fig9}
\end{figure}
\vspace{.5cm}
\begin{figure}[p]
\begin{center}
\leavevmode
\epsfbox{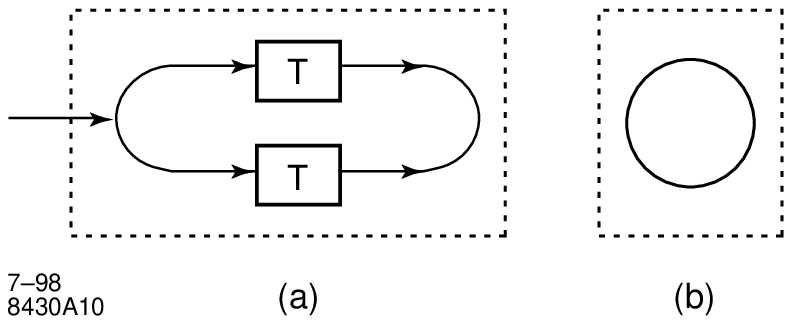}
\end{center}
\label{fig10}
\end{figure}

\end{document}